\documentclass{aastex}
\usepackage{emulateapj5, apjfonts, psfig, epsfig}
\begin{document}

\slugcomment{AJ, in press} 

\title{HST Snapshot Study of Variable Stars In Globular Clusters: \\
Inner Region of NGC~6441\footnotemark}  
\footnotetext{Based on observations with the NASA/ESA {\it Hubble Space 
Telescope}, obtained at the Space Telescope Science Institute, which is 
operated by the Association of Universities for Research in Astronomy, 
Inc., (AURA), under NASA Contract NAS 5-26555.} 

\shorttitle{HST Snapshot: NGC~6441 Variable Stars} 
\shortauthors{Pritzl et al.}

\received{}
\accepted{}
\revised{}

\author{Barton J. Pritzl\altaffilmark{2,3}, Horace A. Smith\altaffilmark{2}, 
Peter B. Stetson\altaffilmark{4}, M\'arcio Catelan\altaffilmark{5}, Allen V. 
Sweigart\altaffilmark{6}, Andrew C. Layden\altaffilmark{7}, R. Michael 
Rich\altaffilmark{8}}

\altaffiltext{2}{Dept. of Physics and Astronomy, Michigan State University, 
East Lansing, MI 48824; pritzl@pa.msu.edu, smith@pa.msu.edu}
\altaffiltext{3}{Current address: National Optical Astronomy Observatories, 
P.O. Box 26732, Tucson, AZ 85726; pritzl@noao.edu}
\altaffiltext{4}{Dominion Astrophysical Observatory, Herzberg Institute 
of Astrophysics, National Research Council, 5071 West Saanich Road, Victoria, 
BC V9E 2E7, Canada; Peter.Stetson@nrc.gc.ca}
\altaffiltext{5}{Pontificia Universidad Cat\'olica de Chile, Departamento 
de Astronom\'\i a y Astrof\'\i sica, Av. Vicu\~{n}a Mackenna 4860, 782-0436 
Macul, Santiago, Chile; mcatelan@astro.puc.cl}
\altaffiltext{6}{NASA Goddard Space Flight Center, Laboratory for Astronomy and 
Solar Physics, Code~681, Greenbelt, MD 20771; sweigart@bach.gsfc.nasa.gov}
\altaffiltext{7}{Department of Physics and Astronomy, 104 Overman Hall, 
Bowling Green State University, Bowling Green, OH 43403; layden@baade.bgsu.edu}
\altaffiltext{8}{Department of Physics and Astronomy, UCLA, 8979 
Math-Sciences Building, Los Angeles, CA 90095-1562; rmn@astro.ucla.edu}

\begin{abstract}

We present the results of a {\it Hubble Space Telescope\/} snapshot 
program to survey the inner region of the metal-rich globular cluster 
NGC~6441 for its variable stars. A total of 57 variable stars was 
found including 38 RR~Lyrae stars, 6 Population~II Cepheids, and 12 long 
period variables. Twenty-four of the RR~Lyrae stars and all of the 
Population~II Cepheids were previously 
undiscovered in ground-based surveys.  Of the RR~Lyrae stars
observed in this survey, 26 are pulsating in the fundamental mode 
with a mean period of 0.753~d and 12 are first-overtone mode 
pulsators with a mean period of 0.365~d.  These values
match up very well with those found in ground-based surveys.  Combining
all the available data for NGC~6441, we find mean periods of
0.759~d and 0.375~d for the RRab and RRc stars, respectively.  We also find 
that the RR~Lyrae in this survey are located in the same regions of a
period-amplitude diagram as those found in ground-based surveys.  The overall 
ratio of RRc to total RR~Lyrae is 0.33. Although NGC~6441 is a metal-rich
globular cluster and would, on that ground, be expected either to have few
RR~Lyrae stars, or to be an Oosterhoff type~I system, its RR~Lyrae more
closely resemble those in Oosterhoff type~II globular clusters. However, even 
compared to typical Oosterhoff type~II systems, the mean period
of its RRab stars is unusually long.  We also derived $I$-band 
period-luminosity relations for the RR~Lyrae stars.

Of the six Population~II Cepheids, five are of W~Virginis type and one 
is a BL~Herculis variable star. This makes NGC~6441, along with NGC~6388, 
the most metal-rich globular cluster known to contain these types of variable
stars. Another variable, V118, may also be a Population~II Cepheid given 
its long period and its separation in magnitude from the RR~Lyrae stars. 
We examine the period-luminosity relation for these Population~II Cepheids 
and compare it to those in other globular clusters and in the Large 
Magellanic Cloud. We argue that there does not appear to be a change 
in the period-luminosity relation slope between the BL~Herculis and 
W~Virginis stars, but that a change of slope does occur when the RV~Tauri
 stars are added to the period-luminosity relation.

\end{abstract}


\section{Introduction}

Globular clusters (GCs) are tracers of the formation 
of the galaxy to which they belong. One key way to analyze the 
stellar populations of a GC is to determine its 
RR~Lyrae (RRL) content, which is tied to the horizontal branch (HB) 
morphology.  Oosterhoff (1939) discovered a fundamental division of Galactic
GCs into two groups based on the mean periods of their 
RRL stars.  It was later shown that this division was associated 
with GC metallicity:  Metal-rich GCs tend to have shorter period RRL 
stars compared to metal-poor GCs (Arp 1955b; Kinman 1959).  Understanding 
the Oosterhoff effect and its effect on RRL stars is key for determining 
accurate distances and ages for the Galactic GCs, and ultimately has 
consequences on how we view the formation of our Galaxy.

\begin{figure*}[t]
  \centerline{\psfig{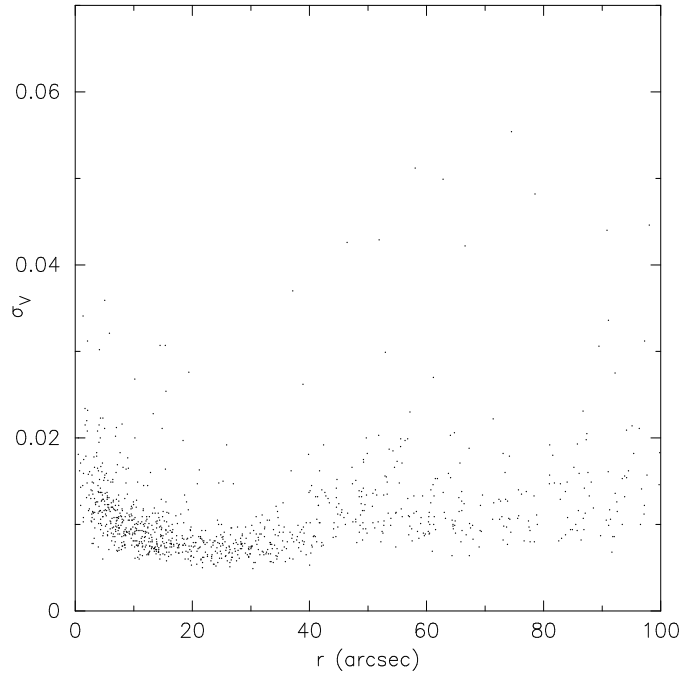}}
  \caption{Photometric uncertainty as a function of clustercentric 
           radius for stars at the horizontal branch level in NGC~6441.  
           The consistency of the photometric uncertainty with radius 
           proves that our errors are not due to crowding, but photon 
           statistics and fitting errors.} 
  \label{Fig01} 
\end{figure*}

Sandage \& Wildey (1967) and
van den Bergh (1967) found that a number of metal-poor GCs
have very different HB morphologies even though they have very similar values
of [Fe/H].\@   It was concluded that at least one parameter,
in addition to metallicity, affects the HB.\@ They
also pointed out that the most distant GCs exhibited strong 
second parameter effects, having red HBs despite their low 
metallicities. Since then it has been shown that a GC's 
galactocentric distance plays a role in the strength of the 
second parameter in the sense that at small galactocentric 
distances the HB morphology and [Fe/H] are closely tied to one 
another, while going out to larger distances one finds that there 
is more scatter in the HB morphologies, with the HB type becoming
redder for a given [Fe/H] (Searle \& Zinn 1978; Zinn 1980; 
Lee, Demarque, \& Zinn 1994).  A difference in cluster ages has
been proposed as one explanation for the second parameter effect, but 
there are cases where this may not be the only explanation for unusual 
HB morphologies (e.g., Fusi Pecci, Bellazzini, \& Ferraro 1996; Fusi 
Pecci \& Bellazzini 1997).  Any second-parameter affecting the HB 
morphology will also affect any RRL stars found in that cluster.  
Thus, as pointed out by Catelan, Sweigart, \& Borissova (1998), RRL stars 
may also provide a probe into the second-parameter phenomenon, which 
in turn may be directly related to the origin of the Oosterhoff 
dichotomy itself (Castellani 1983; Renzini 1983; Lee \& Zinn 1990).

While the second-parameter effect has primarily been found 
in more metal-poor GCs, two metal-rich systems, NGC~6388 
and NGC~6441, have been found to have very unusual HB morphologies for 
their metallicities (Rich et al.\ 1997).  With metallicities of 
${\rm [Fe/H]} \sim -0.6$ these clusters not only have strong red HBs 
similar to the canonical metal-rich GC 47~Tucanae, they also 
have components to their HBs that extend blueward across the instability 
strip and terminate in blue tails.  In addition, the red HBs in both 
clusters are sloped toward increasing brightness as one moves blueward 
along the HB, which cannot be explained by differential reddening alone.  
To make these GCs even more interesting, Sweigart \& Catelan 
(1998) have claimed that non-canonical effects are needed to explain 
the sloped nature of the HBs.  Rich et al.\ speculated that the 
high central concentrations of NGC~6388 and NGC~6441 may have led to 
higher interactions in the inner regions of the clusters, resulting in 
the bluer HBs---without however providing an explanation for the sloping 
nature of the HBs.  However, both Rich et al.\ and Layden et al.\ 
(1999) found that this hypothesis did not hold because the blue HB stars 
were not centrally concentrated.

In attempting to model the HBs in NGC~6388 and NGC~6441, Sweigart \& Catelan
(1998) proposed three alternatives, (i) a high cluster helium abundance,
(ii) an increase in helium core mass due to rotation on the red giant
branch, and (iii) enhancement of the envelope helium abundance due to deep
mixing. The high helium abundance scenario has subsequently been ruled
out on observational grounds (Layden et al.\ 1999; Raimondo et al.\ 2002).
Other explanations for these unusual HB morphologies have ranged from 
increased mass loss along the red giant branch, to a spread in metallicity, 
to the clusters being dwarf galaxy remnants (Piotto et al.\ 1997; 
Sweigart 2002; Ree et al.\ 2002).

All three scenarios introduced by Sweigart \& Catelan (1998) require 
that the HBs of NGC~6388 and NGC~6441 be unusually bright for the metal 
abundances of the clusters. As a consequence, any RRL stars found in 
these clusters would also be unusually bright and would therefore have 
unusually long periods for their metallicity. Surveys
of the variable star content of NGC~6388 (Silbermann et al.\ 1994; Pritzl 
et al.\ 2002) and NGC~6441 (Layden et al.\ 1999; Pritzl et al.\ 2001) 
found that their RRL stars did indeed have periods that are unusually 
long for the cluster metallicity.  On the other hand, the gravities 
(for the hot HB stars) measured by Moehler, Sweigart, \& Catelan (1999) 
imply that the blue HB stars in these clusters are 
not anomalously bright, contrary to the Sweigart \& Catelan (1998) 
predictions.

Due to the compact nature of these clusters [$\log\,\rho \,\ (L_{\odot} 
\,\ {\rm pc}^{-3}) = 5.34$ (NGC~6388), 5.25 (NGC~6441)], it is very difficult 
to survey the inner regions of the clusters, where most of the 
cluster variables are expected to be found, from the ground.  Making use 
of the {\it Hubble Space Telescope\/} (HST) snapshot program, we have 
obtained images of the inner region of NGC~6441 to survey its variable 
star content with the Wide Field Planetary Camera~2 (WFPC2). We present 
in this paper the results of our survey and compare the properties of the
variable stars found in the inner region of the cluster to those variables 
found in the outer regions from the ground-based surveys. We also
examine the nature of the Population~II Cepheids (P2Cs) found in
NGC~6441 and NGC~6388 and compare them with those in other GCs.

\section{Observations and Reductions}

\begin{figure*}[t] 
  \figurenum{2a}
  \centerline{\psfig{figure=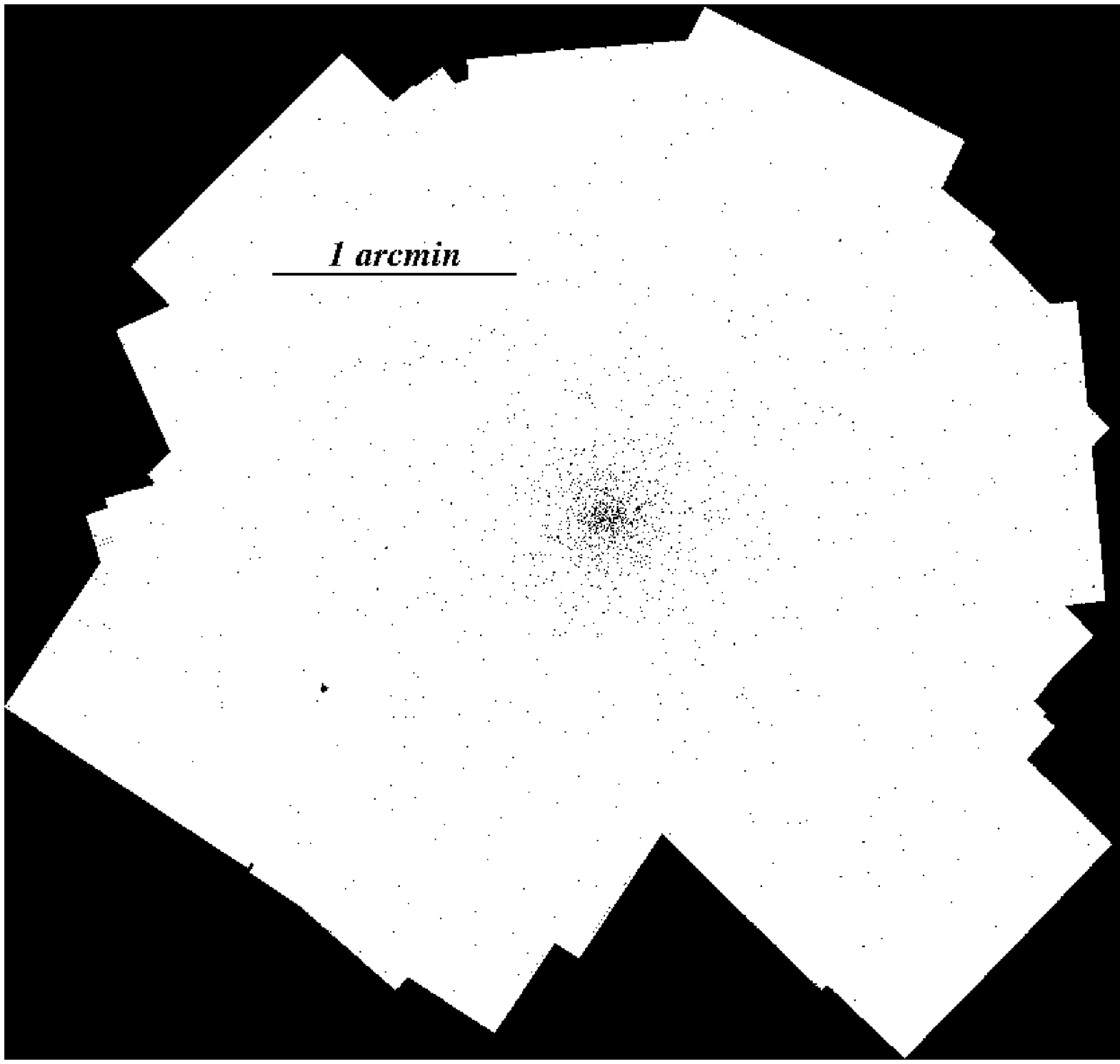,height=3.25in,width=3.50in}} 
  \caption{NGC~6441 finding charts marking all of the variable stars 
           found in this paper.  Figure~2a shows the full field-of-view 
           for our observations.  The image is 283-by-267 arcseconds at 
           maximum extent, with a scale of 22 pixels per arcsecond.  
           North is up and east is to the left.} 
  \label{Fig02a}
\end{figure*} 

\begin{figure*}[t] 
  \figurenum{2b}
  \centerline{\psfig{figure=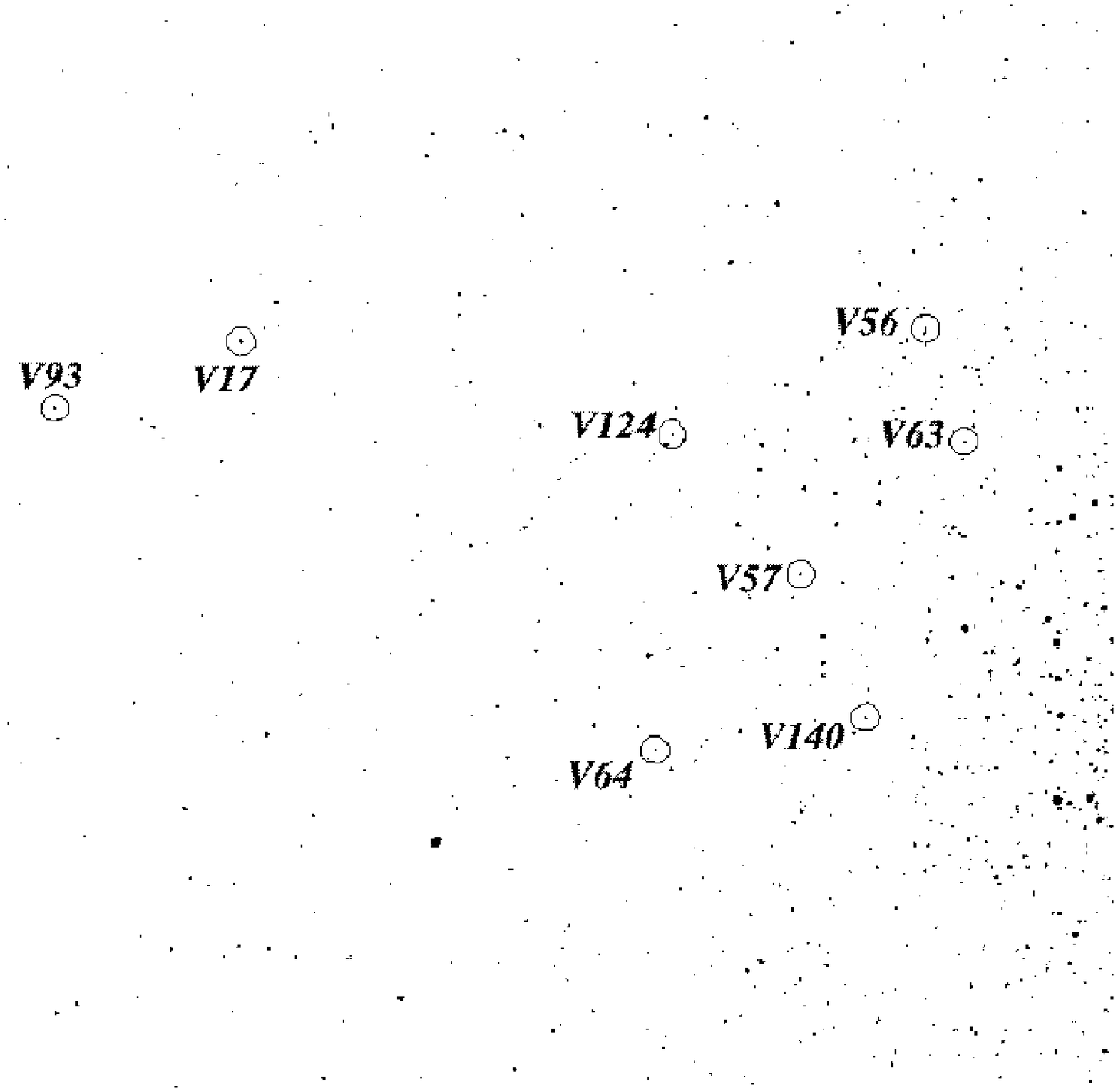,height=3.25in,width=3.50in}} 
  \caption{NGC~6441 finding charts marking all of the variable stars 
           found in this paper.  Figure~2a shows the full field-of-view 
           for our observations.  The image is 283-by-267 arcseconds at 
           maximum extent, with a scale of 22 pixels per arcsecond.  
           North is up and east is to the left.} 
  \label{Fig02b}
\end{figure*} 

\begin{figure*}[t] 
  \figurenum{2c}
  \centerline{\psfig{figure=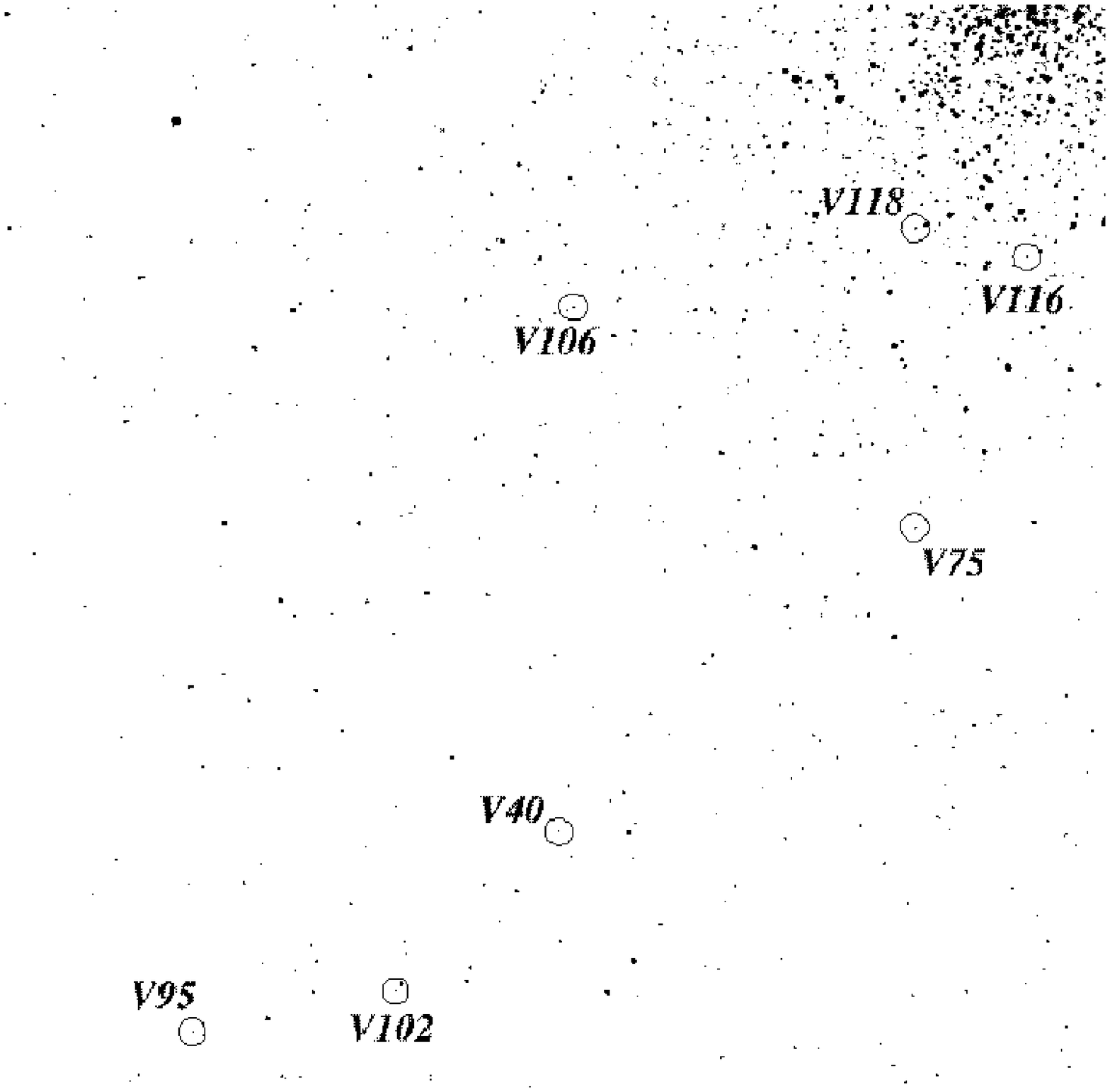,height=3.25in,width=3.50in}} 
  \caption{NGC~6441 finding charts marking all of the variable stars 
           found in this paper.  Figure~2a shows the full field-of-view 
           for our observations.  The image is 283-by-267 arcseconds at 
           maximum extent, with a scale of 22 pixels per arcsecond.  
           North is up and east is to the left.} 
  \label{Fig02c}
\end{figure*} 

\begin{figure*}[t] 
  \figurenum{2d}
  \centerline{\psfig{figure=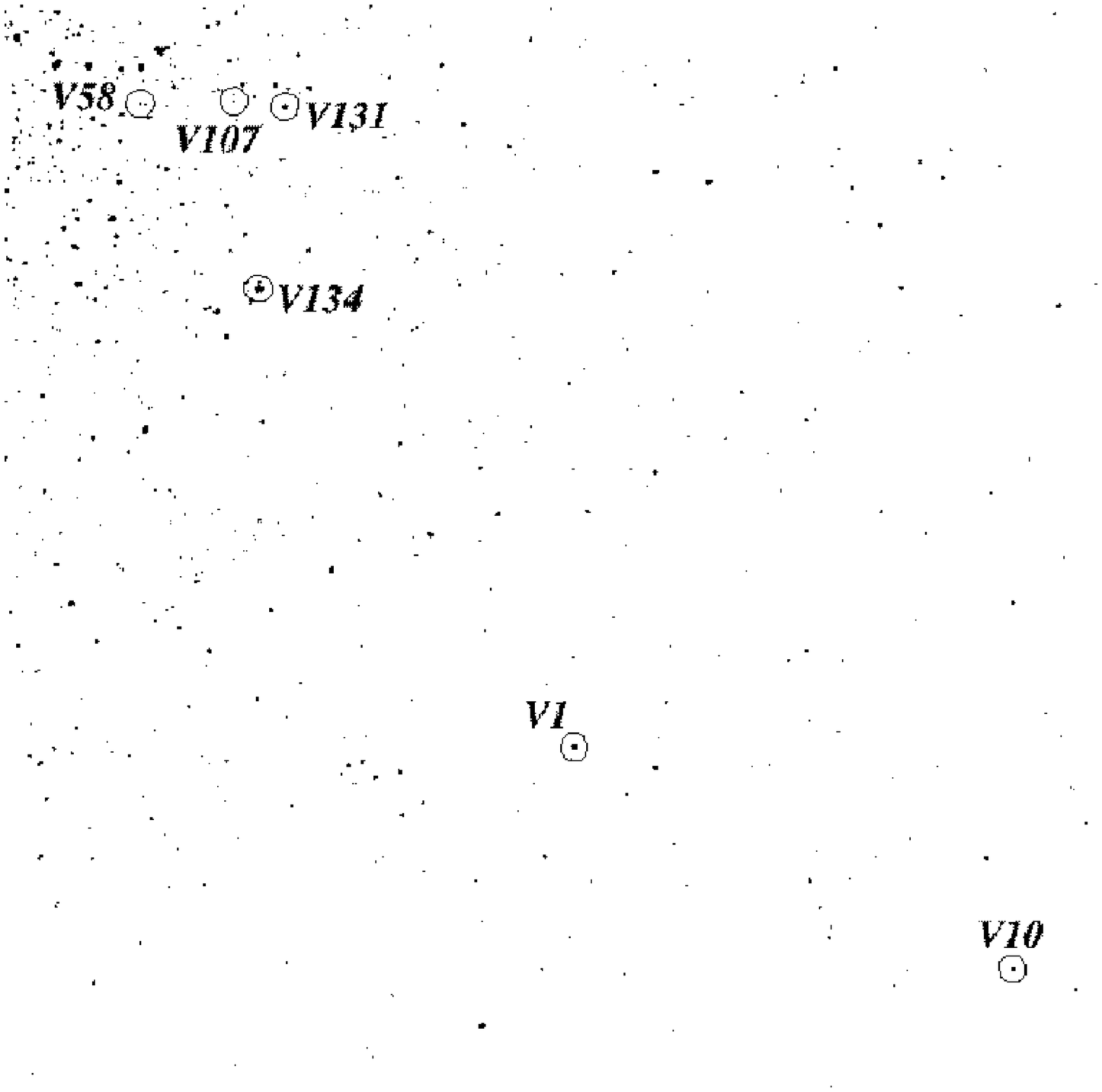,height=3.25in,width=3.50in}} 
  \caption{NGC~6441 finding charts marking all of the variable stars 
           found in this paper.  Figure~2a shows the full field-of-view 
           for our observations.  The image is 283-by-267 arcseconds at 
           maximum extent, with a scale of 22 pixels per arcsecond.  
           North is up and east is to the left.} 
  \label{Fig02d}
\end{figure*} 

\begin{figure*}[t] 
  \figurenum{2e}
  \centerline{\psfig{figure=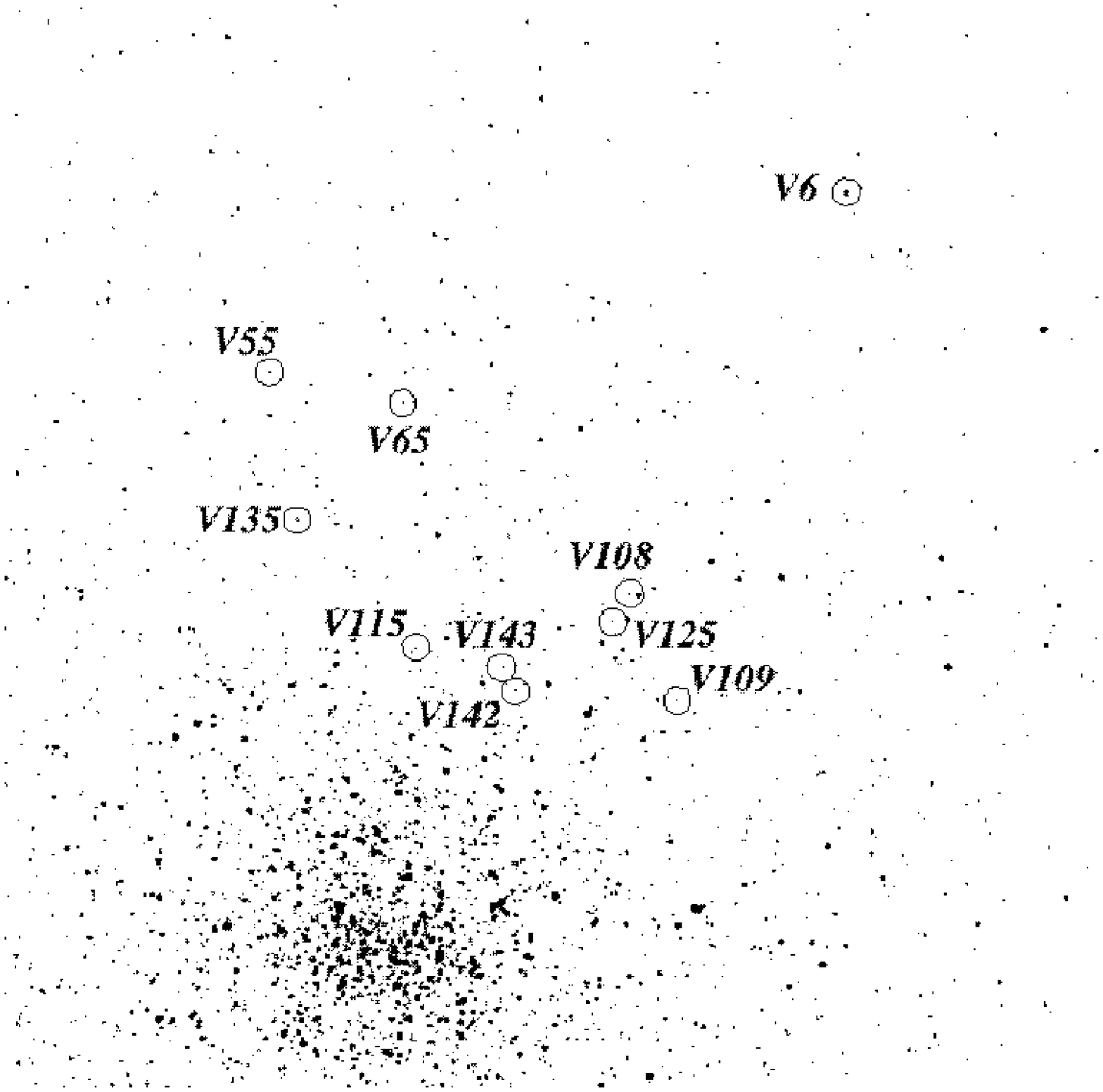,height=3.25in,width=3.50in}} 
  \caption{NGC~6441 finding charts marking all of the variable stars 
           found in this paper.  Figure~2a shows the full field-of-view 
           for our observations.  The image is 283-by-267 arcseconds at 
           maximum extent, with a scale of 22 pixels per arcsecond.  
           North is up and east is to the left.} 
  \label{Fig02e}
\end{figure*} 

\begin{figure*}[t] 
  \figurenum{2f}
  \centerline{\psfig{figure=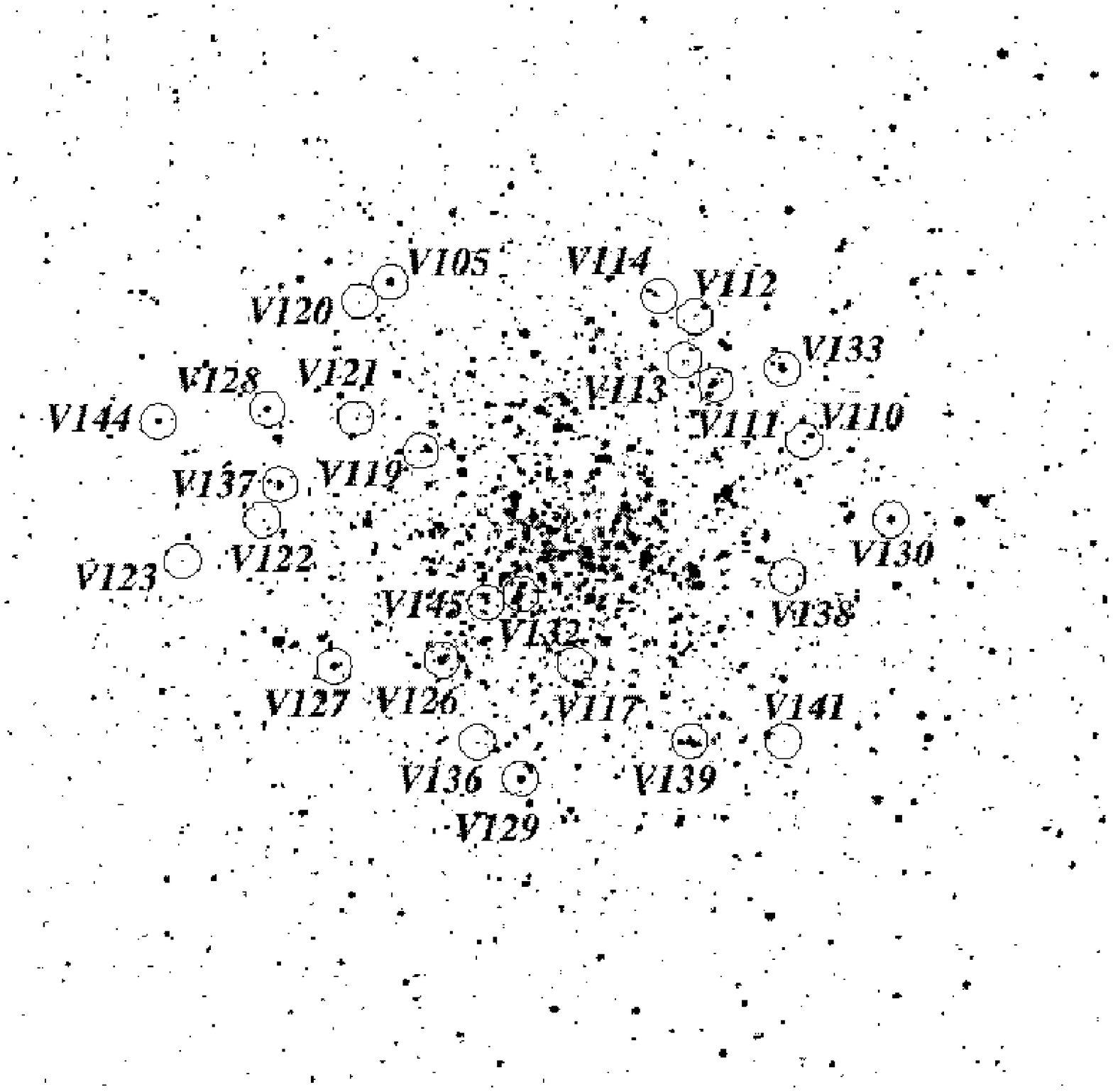,height=3.25in,width=3.50in}} 
  \caption{NGC~6441 finding charts marking all of the variable stars 
           found in this paper.  Figure~2a shows the full field-of-view 
           for our observations.  The image is 283-by-267 arcseconds at 
           maximum extent, with a scale of 22 pixels per arcsecond.  
           North is up and east is to the left.} 
  \label{Fig02f}
\end{figure*}

Eighteen sets of observations in $B$, $V$, and $I$ were taken by the 
HST with the Wide-Field Planetary Camera~2 starting on 1999 March 16 
and ending on 2000 June 30 as part of the SNAP~8251 program. The complete 
list of observations is given in Table~1, including NGC~6441 HST archive
observations which were added to our reductions (GO~5667, GO~6095,
and GO~6780). Exposure times for each filter are also listed in
Table~1. NGC~6441 was centered on the WFC3 chip for our program.
No restrictions were made on the orientation of the observations.
Therefore, rotations between the images, in addition to translations,
produced partial and varying overlap among the images obtained on 
the other three chips.

The observations were reduced using the {\sc daophot/ allstar/ allframe} 
routines (e.g., Stetson 1987, 1994). We used standard point-spread 
functions that had been prepared for each chip and filter in
the course of the {\it Hubble Space Telescope Key Project on the
Extragalactic Distance Scale\/} (e.g., Freedman et al.\ 2001; Stetson
et al.\ 1998) and our previous research on outer-halo globulars
(e.g., Harris et al.\ 1997, Stetson et al.\ 1999). After the geometric
relationships relating the various images to a common coordinate system
and a master star list for the field had been generated, {\sc allframe} was
used to obtain final stellar positions on the sky, and a magnitude
measurement for each star at each epoch. Aperture corrections
were determined from comparatively bright, isolated stars in each image
to relate them to a well-defined, common instrumental zero-point. 
Charge-transfer efficiency corrections and fundamental photometric
zero-points similar to those of Stetson (1998), but based on a much
larger database, and the color corrections of Holtzman et al.\
(1995) were applied in the standard fashion. 

Figure~1 shows the photometric uncertainty in $V$ at the level of the 
HB as a function of clustercentric radius.  We assumed $m_{V, {\rm RR}} 
= 17.54$~mag and chose a range of magnitudes around this value of 
$\pm 0.25$~mag.  All candidate variable stars were removed.  The $\sigma_V$ 
relationship is nearly independent of radius.  This suggests that at 
these limits our photometric errors are dominated by photon 
statistics and fitting errors and not by crowding for most stars.  The 
increased photometric uncertainty at large radii occurs because, with 
the differing roll angles of the spacecraft, not all stars could be 
measured at all epochs.

Variable stars were found by a variety of methods including a variant of
the Welch \& Stetson (1993) algorithm (see Stetson 1996).  Finding charts 
for the variables can be seen in Figure~2.  We used the period finding 
routines described in Layden et al.\ (1999) and Layden \& Sarajedini 
(2000) to determine the periods of the variable stars.  This technique 
involves folding each star's magnitude-time data by a sequence of periods, 
and at each period, fitting the resulting light curve with a set of template 
light curves.  If the period being tested is near the star's true period, 
the template will fit the data well and produce a small value of 
$\chi^2$.  Probable periods are represented by $\chi^2$ minima in the 
array of periods and templates considered.  We searched for periods in 
the range appropriate for each class of variable; e.g., for RRL stars 
the range was 0.2 to 1.0~day.  Once a period was found, the search was 
narrowed to about 0.0001~day around the candidate period to determine 
the best period.  In almost all cases, we found a single, dominant 
$\chi^2$ minimum, suggesting that aliasing was not a problem despite the 
irregular sampling and long time interval over which the data were gathered.  
The few exceptions involved a small number of first-overtone (RRc) stars 
which produced light curves with significant scatter, and long period 
variables (LPVs) whose entire light cycle was not covered by our 
observations.  We derived pulsation amplitudes and intensity-weighted 
magnitudes from the fitted templates.  We estimate the uncertainty in 
these magnitudes to be 0.02--0.03~mag for the RRL stars and Population~II 
Cepheid (P2C) stars.  Typical uncertainties in the RRL and P2C periods 
are estimated to be 0.00002 and 0.002~days.  The magnitude and period 
uncertainties are considerably larger for the LPV stars.  In part, this is 
because the templates we used are not optimal for LPV stars, but a more 
significant discrepancy arises because many of the LPVs may not 
demonstrate pulsations with a regular period.  The analysis here is 
intended only to classify the type of pulsation and to establish 
approximate mean-light parameters so the stars may be placed in the CMD.\@  
A comprehensive analysis of the LPV stars in NGC~6441 is reserved for a 
future paper.

\begin{figure*}[t]
  \figurenum{3}
  \centerline{\psfig{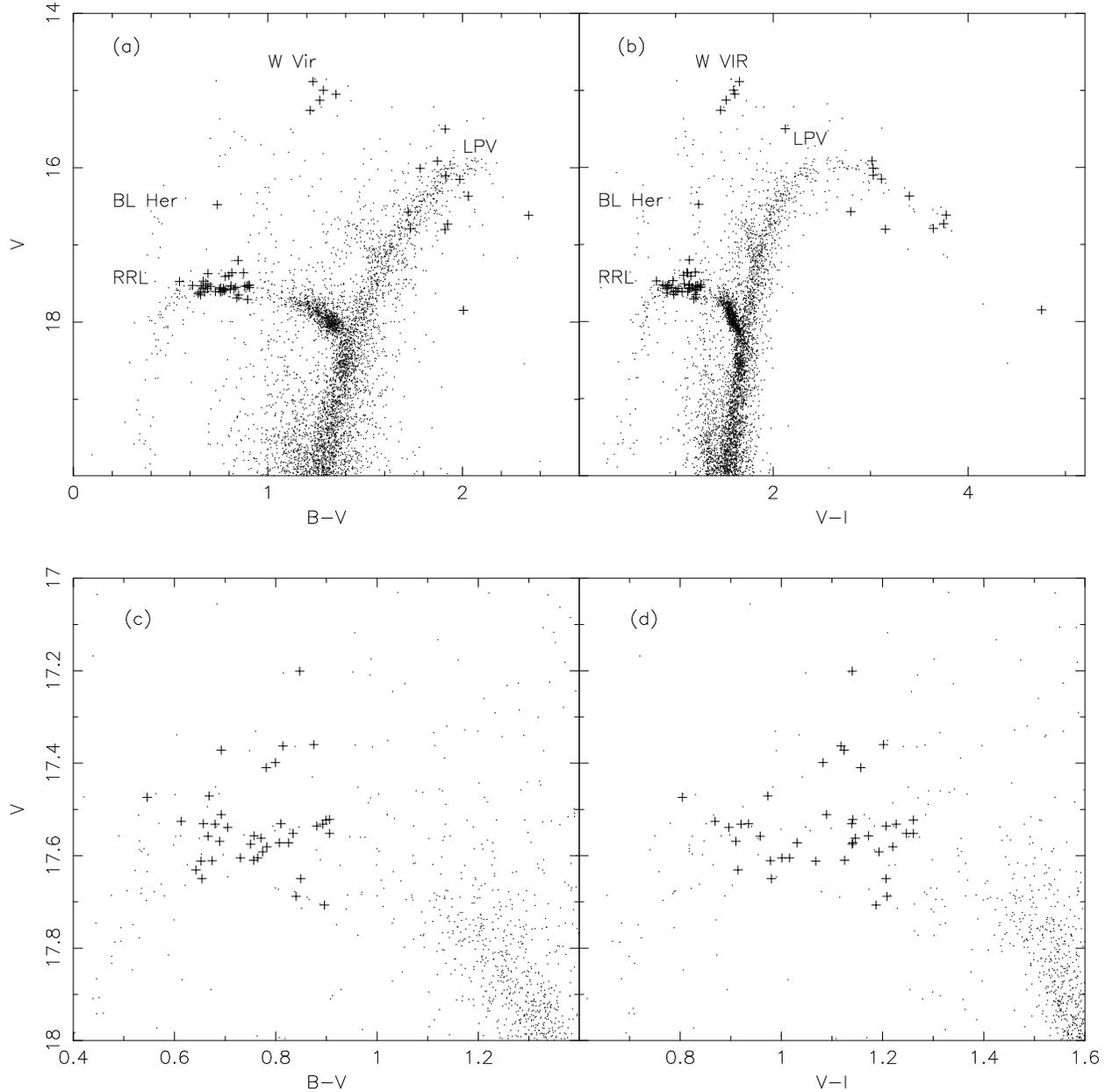}}
  \caption{NGC~6441 color-magnitude diagram showing the location of the 
           variable stars (plusses) found in this survey.  The top two 
           plots (a \& b) show those found along the horizontal branch 
           are clearly the RR~Lyrae stars, while those found brighter 
           than the horizontal branch are the Population~II Cepheids.  
           Those variables along the red giant branch are the long period 
           variables. As noted in the text, these have uncertain magnitudes 
           due to the method by which we calculated them.  In the bottom 
           two plots (c \& d) we show a close-up of the horizontal branch.  
           Here we see a number of candidate RR~Lyrae stars are brighter 
           than the majority of the RR~Lyrae stars along the horizontal 
           branch.  These stars may be evolved RR~Lyrae stars, while 
           the brightest of these, V118, may be a Population~II Cepheid.} 
  \label{Fig03} 
\end{figure*}

Figure~3 shows the positions of the variables in the
color-magnitude diagrams for NGC~6441 in $V$ versus (a) $\bv$ and
(b) $V-I$, where variables are plotted according to their 
intensity-weighted $\langle V \rangle$ magnitudes and $\langle B \rangle - 
\langle V \rangle$ and $\langle V \rangle - \langle I \rangle$ colors.  
Figures~3c and 3d give an expanded view of the HB.  The variable stars 
along the HB are clearly RRL stars; those brighter 
than the HB and located toward the red but blueward of the red giant branch
are P2Cs; those found after the turnover of the red giant branch are 
LPVs.  However, as noted above, a number of the LPVs have poor periods 
due to aliasing and have larger uncertainties in their mean magnitudes.  
We also found one $\delta$~Scuti star (V95), but given its magnitude 
it is likely a member of the field.  Due to the small number of data points and 
the gaps in the phase coverage, we did not attempt to do a Fourier 
analysis of the light curves.

In the next sections we present the detailed results of our survey for
variable stars and discuss their properties. Comparisons are also made
to the variables found in ground-based
surveys to determine any similarities and differences in their properties
as compared to those found in this study.

\section{RR~Lyrae Stars}

\begin{figure*}[t]
  \figurenum{4}
  \centerline{\psfig{figure=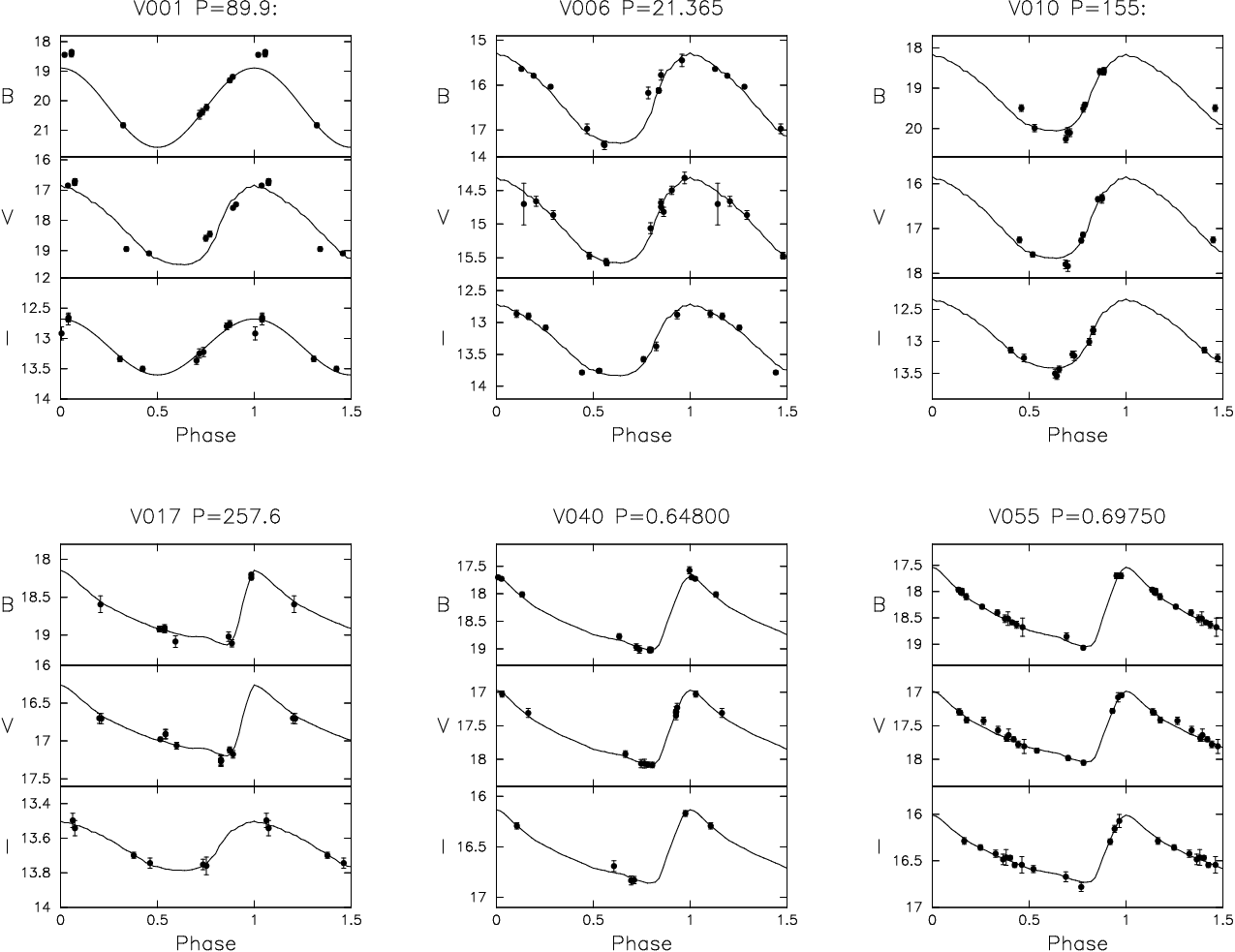,height=5.5in,width=7.0in,angle=-90}} 
  \caption{NGC~6441 light curves in $B$, $V$, and $I$.  Lines shown are 
           template fits to the data.  In some cases, mostly with the 
           $I$ data, the template was not able to find a good fit due 
           to the quality of the data. Long period variables do not have 
           good fits since the templates were not created to deal with 
           this type of variable.} 
  \label{Fig04}
\end{figure*}

\begin{figure*}[t]
  \figurenum{4 cont}
  \centerline{\psfig{figure=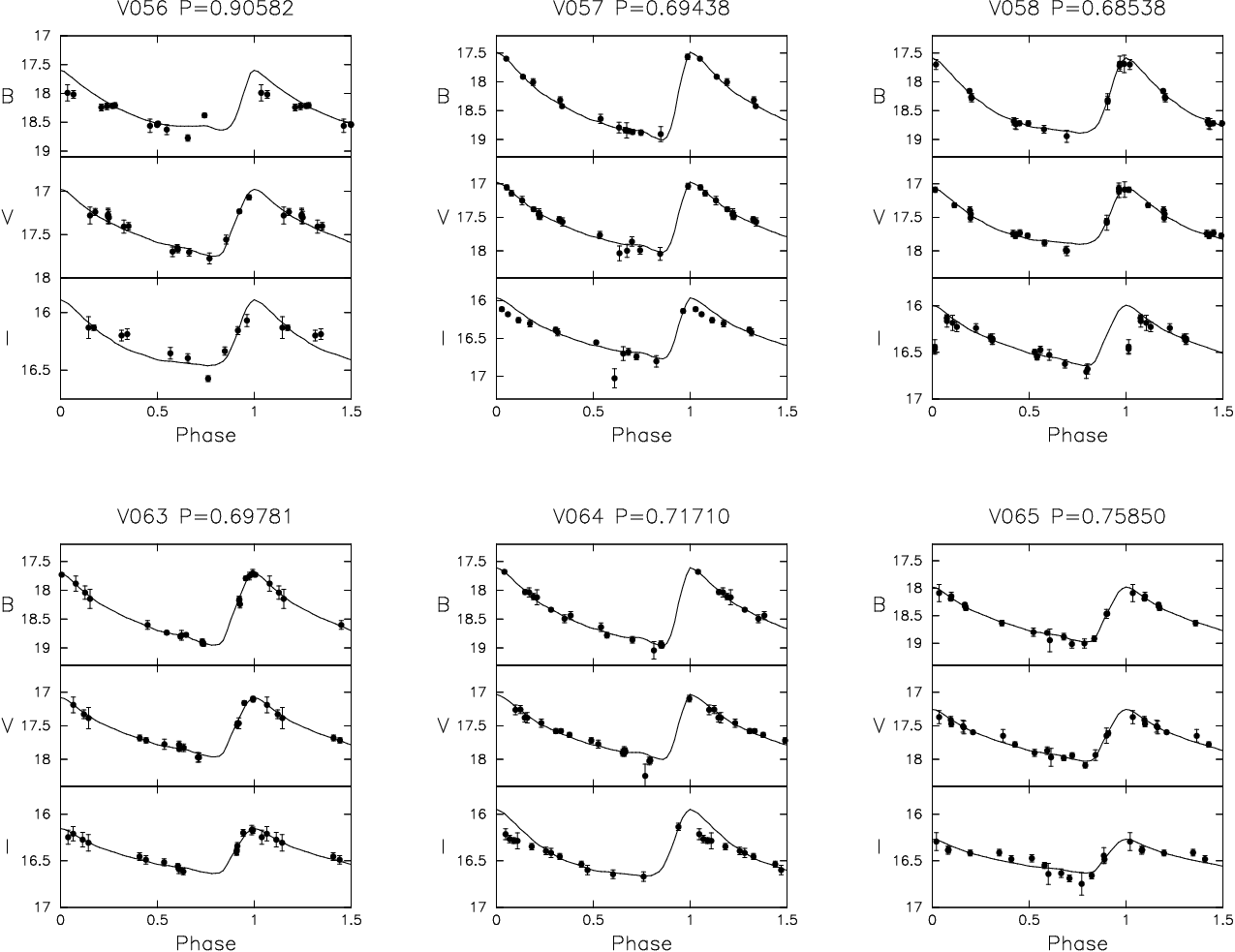,height=5.5in,width=7.0in,angle=-90}} 
  \caption{NGC~6441 light curves in $B$, $V$, and $I$.  Lines shown are 
           template fits to the data.  In some cases, mostly with the 
           $I$ data, the template was not able to find a good fit due 
           to the quality of the data. Long period variables do not have 
           good fits since the templates were not created to deal with 
           this type of variable.} 
  \label{Fig04}
\end{figure*}

\begin{figure*}[t]
  \figurenum{4 cont}
  \centerline{\psfig{figure=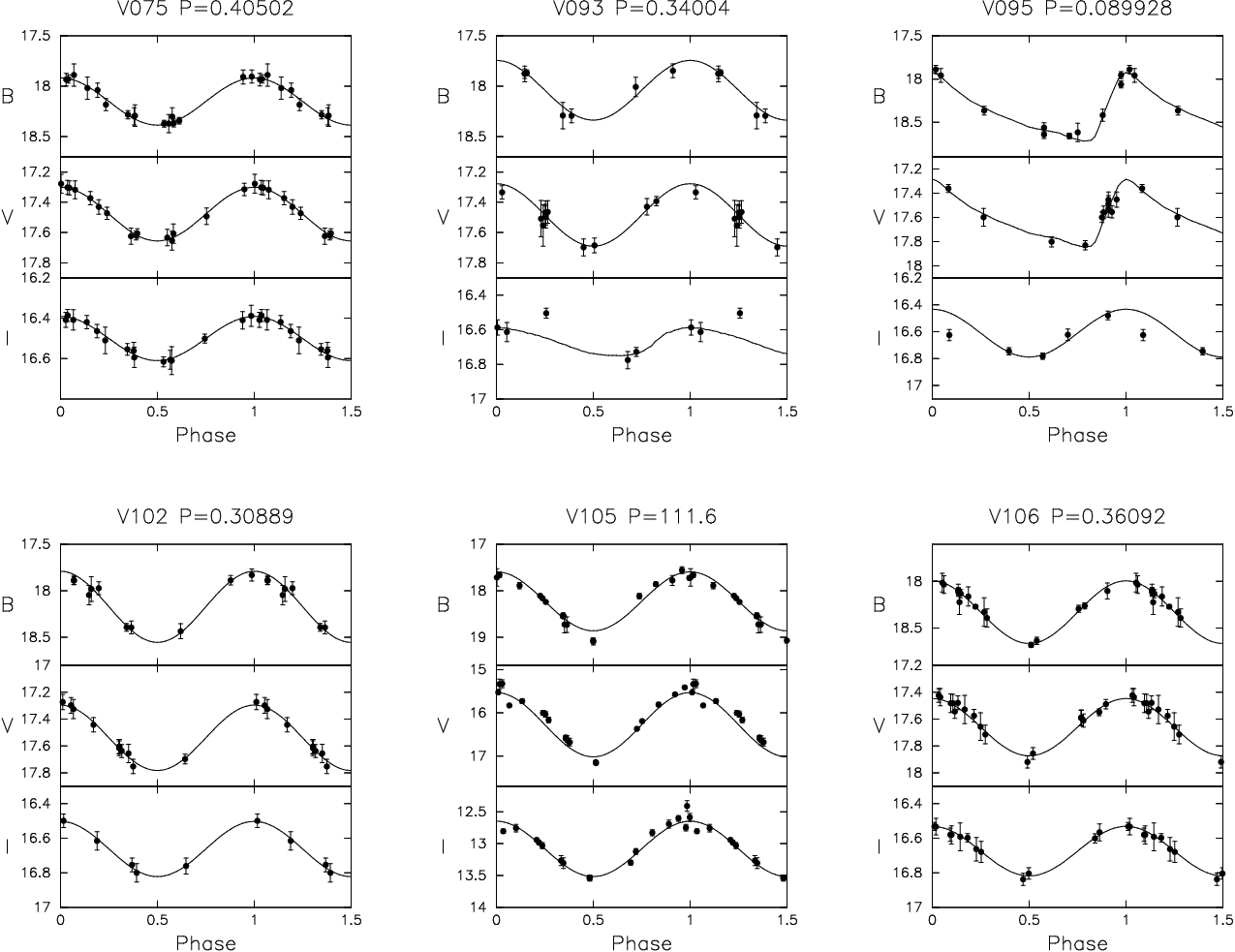,height=5.5in,width=7.0in,angle=-90}} 
  \caption{NGC~6441 light curves in $B$, $V$, and $I$.  Lines shown are 
           template fits to the data.  In some cases, mostly with the 
           $I$ data, the template was not able to find a good fit due 
           to the quality of the data. Long period variables do not have 
           good fits since the templates were not created to deal with 
           this type of variable.} 
  \label{Fig04}
\end{figure*}

\begin{figure*}[t]
  \figurenum{4 cont}
  \centerline{\psfig{figure=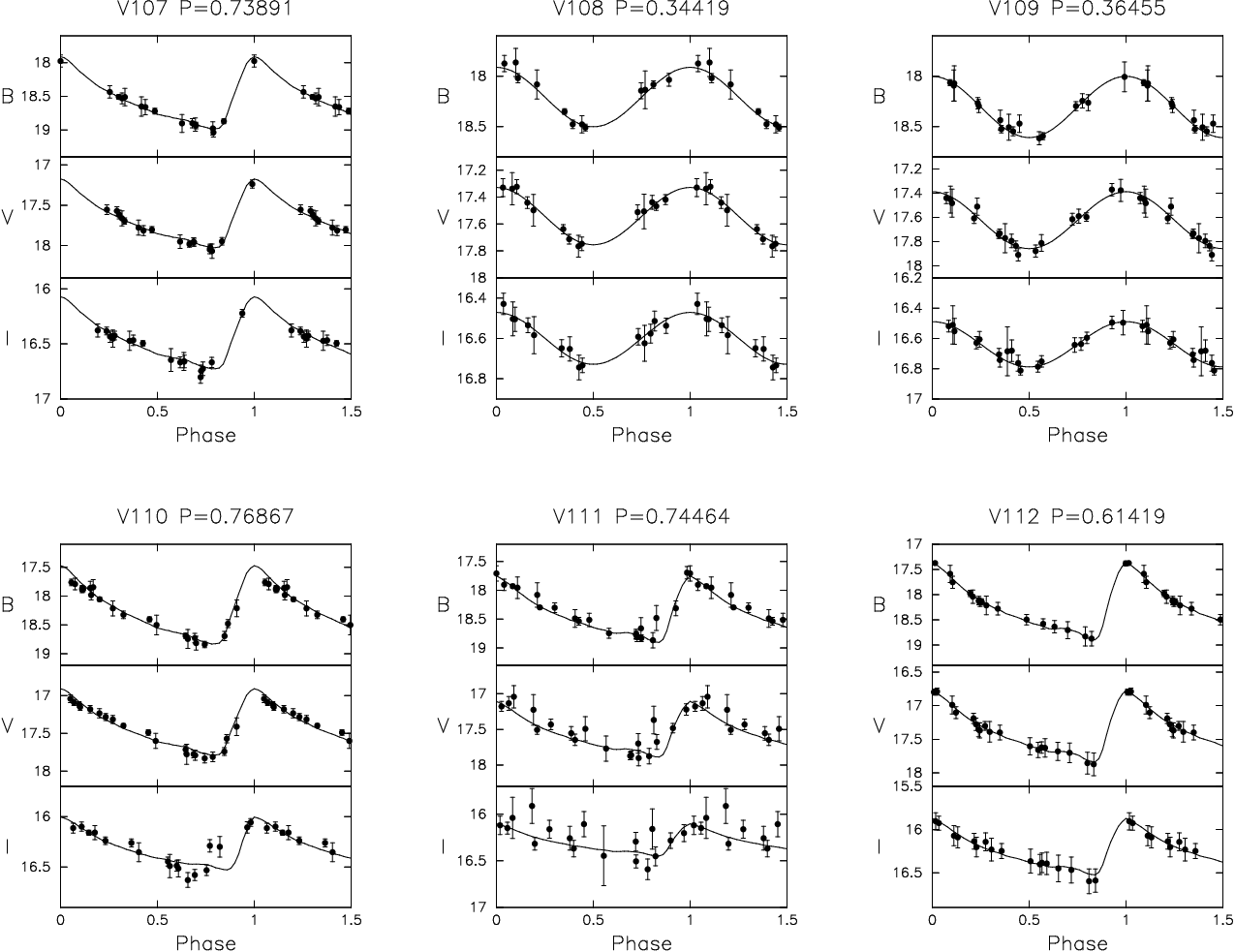,height=5.5in,width=7.0in,angle=-90}} 
  \caption{NGC~6441 light curves in $B$, $V$, and $I$.  Lines shown are 
           template fits to the data.  In some cases, mostly with the 
           $I$ data, the template was not able to find a good fit due 
           to the quality of the data. Long period variables do not have 
           good fits since the templates were not created to deal with 
           this type of variable.} 
  \label{Fig04}
\end{figure*}

\begin{figure*}[t]
  \figurenum{4 cont}
  \centerline{\psfig{figure=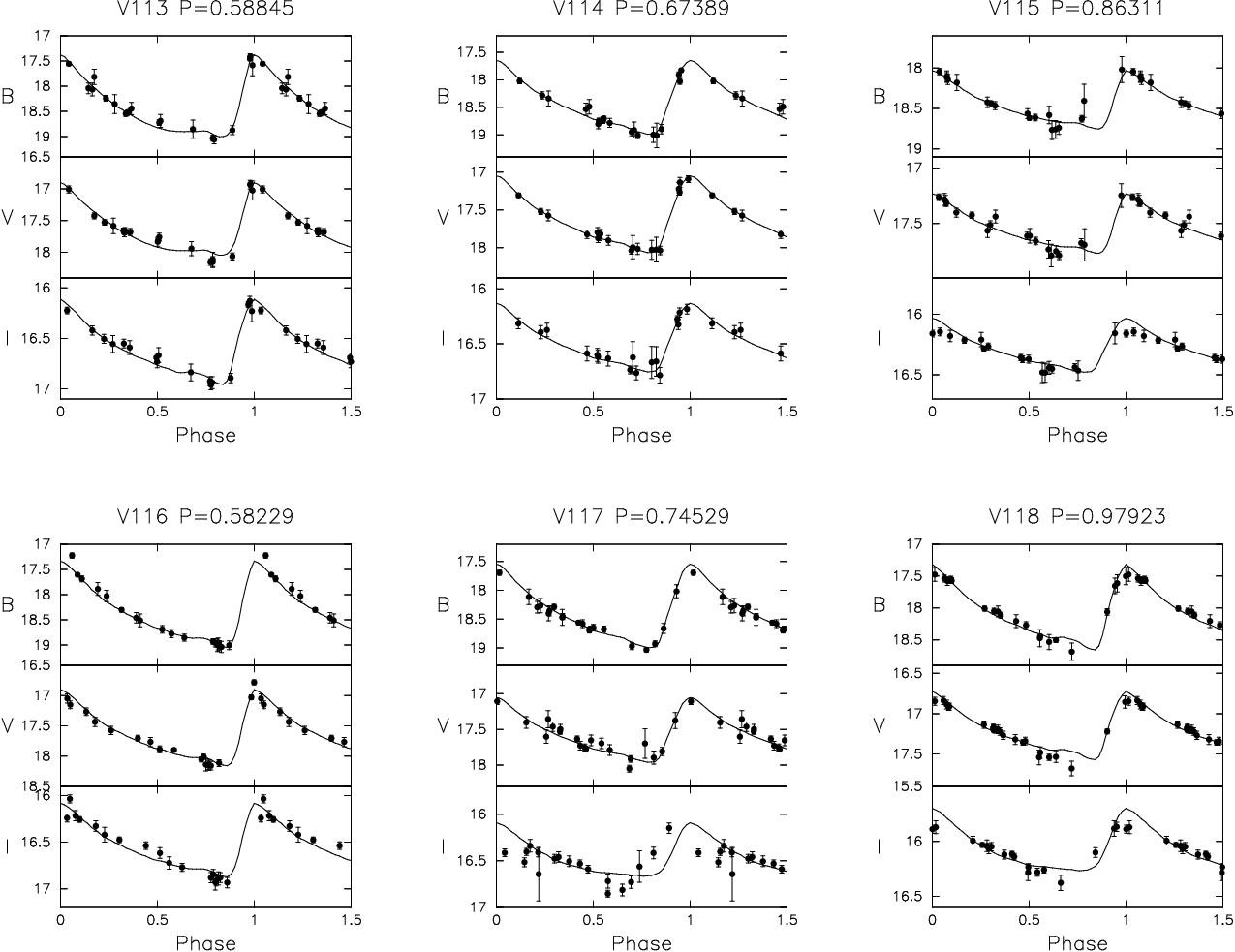,height=5.5in,width=7.0in,angle=-90}} 
  \caption{NGC~6441 light curves in $B$, $V$, and $I$.  Lines shown are 
           template fits to the data.  In some cases, mostly with the 
           $I$ data, the template was not able to find a good fit due 
           to the quality of the data. Long period variables do not have 
           good fits since the templates were not created to deal with 
           this type of variable.} 
  \label{Fig04}
\end{figure*}

\begin{figure*}[t]
  \figurenum{4 cont}
  \centerline{\psfig{figure=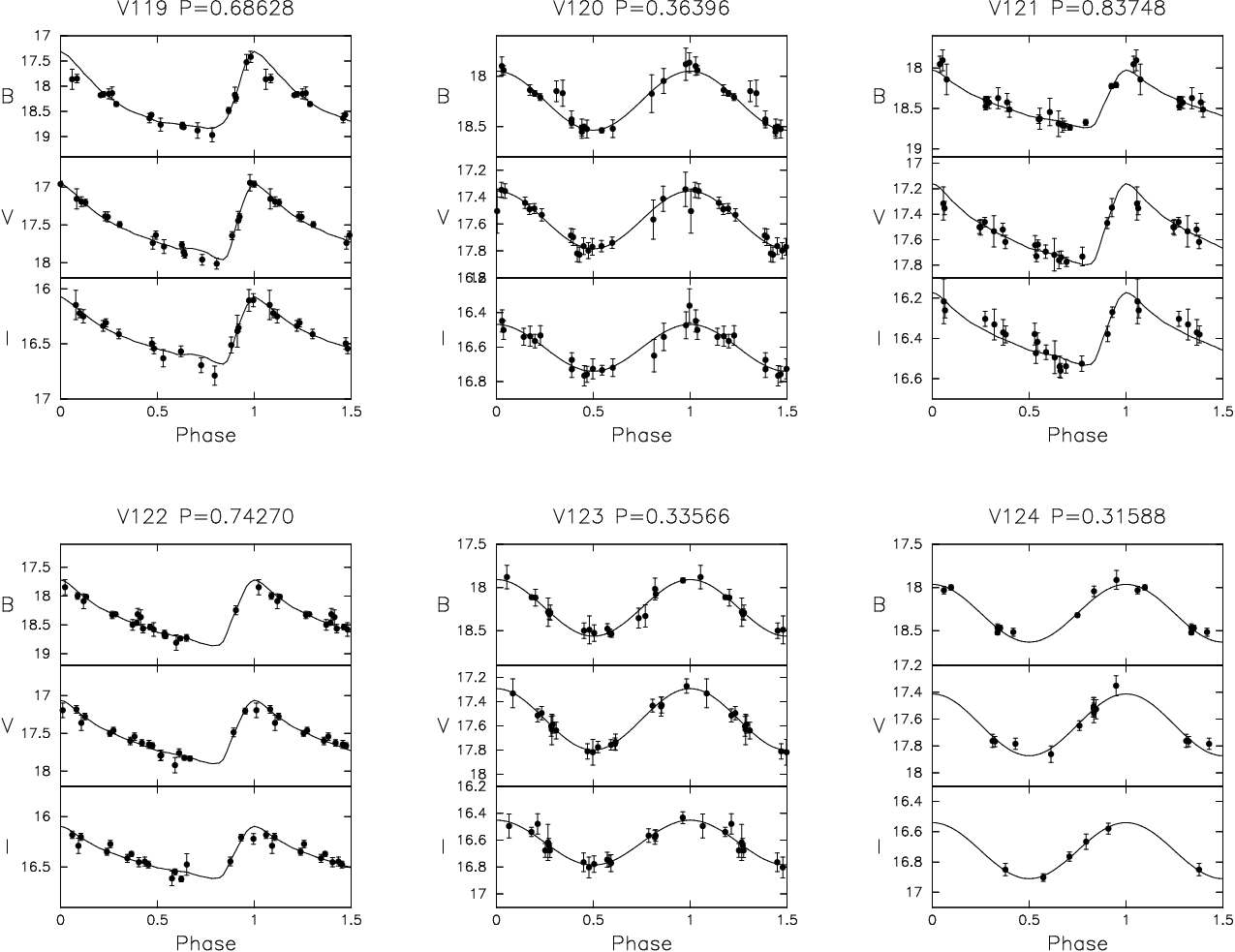,height=5.5in,width=7.0in,angle=-90}} 
  \caption{NGC~6441 light curves in $B$, $V$, and $I$.  Lines shown are 
           template fits to the data.  In some cases, mostly with the 
           $I$ data, the template was not able to find a good fit due 
           to the quality of the data. Long period variables do not have 
           good fits since the templates were not created to deal with 
           this type of variable.} 
  \label{Fig04}
\end{figure*}

\begin{figure*}[t]
  \figurenum{4 cont}
  \centerline{\psfig{figure=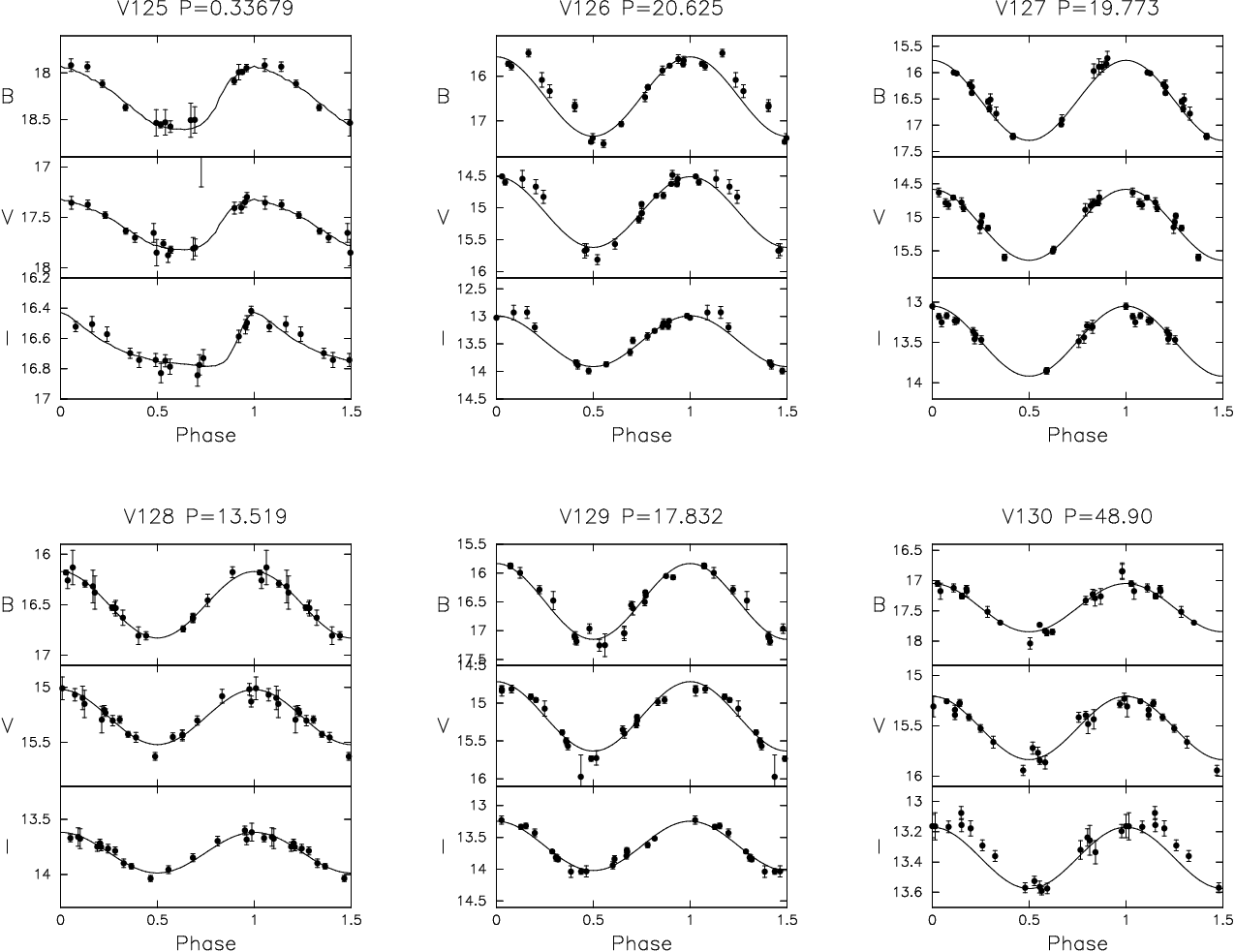,height=5.5in,width=7.0in,angle=-90}} 
  \caption{NGC~6441 light curves in $B$, $V$, and $I$.  Lines shown are 
           template fits to the data.  In some cases, mostly with the 
           $I$ data, the template was not able to find a good fit due 
           to the quality of the data. Long period variables do not have 
           good fits since the templates were not created to deal with 
           this type of variable.} 
  \label{Fig04}
\end{figure*}

\begin{figure*}[t]
  \figurenum{4 cont}
  \centerline{\psfig{figure=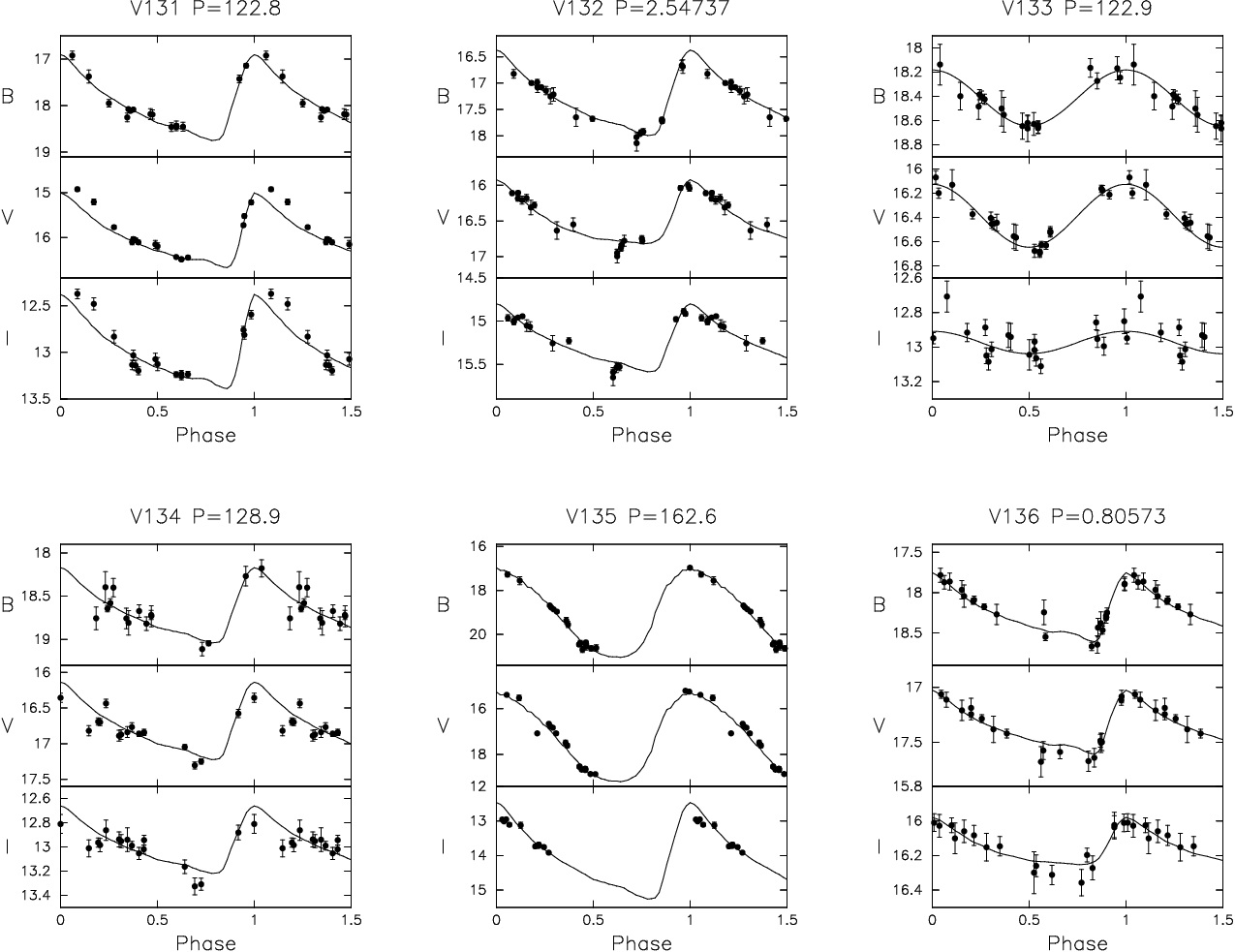,height=5.5in,width=7.0in,angle=-90}} 
  \caption{NGC~6441 light curves in $B$, $V$, and $I$.  Lines shown are 
           template fits to the data.  In some cases, mostly with the 
           $I$ data, the template was not able to find a good fit due 
           to the quality of the data. Long period variables do not have 
           good fits since the templates were not created to deal with 
           this type of variable.} 
  \label{Fig04}
\end{figure*}

\begin{figure*}[t]
  \figurenum{4 cont}
  \centerline{\psfig{figure=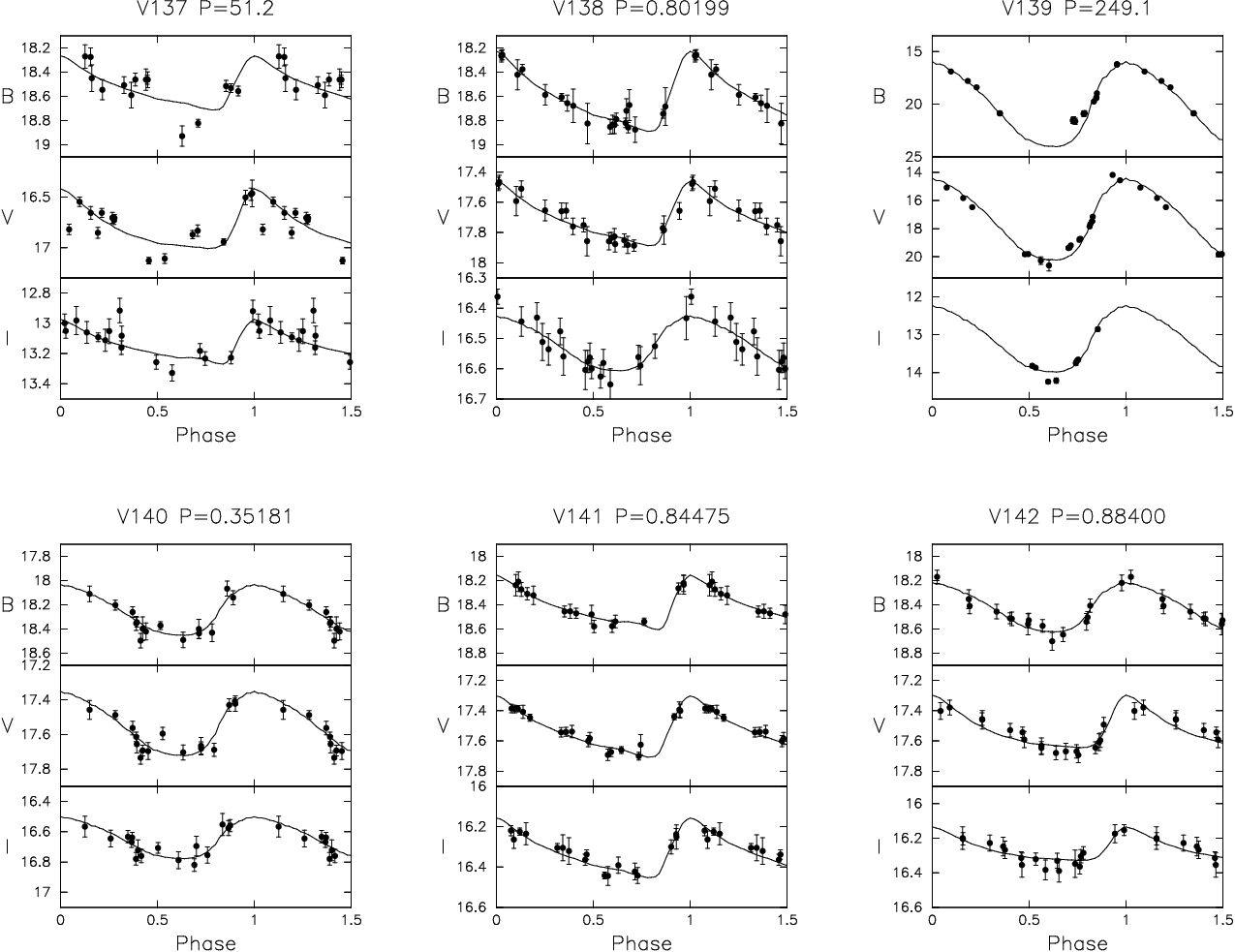,height=5.5in,width=7.0in,angle=-90}} 
  \caption{NGC~6441 light curves in $B$, $V$, and $I$.  Lines shown are 
           template fits to the data.  In some cases, mostly with the 
           $I$ data, the template was not able to find a good fit due 
           to the quality of the data. Long period variables do not have 
           good fits since the templates were not created to deal with 
           this type of variable.} 
  \label{Fig04}
\end{figure*}

\begin{figure*}[t]
  \figurenum{4 cont}
  \centerline{\psfig{figure=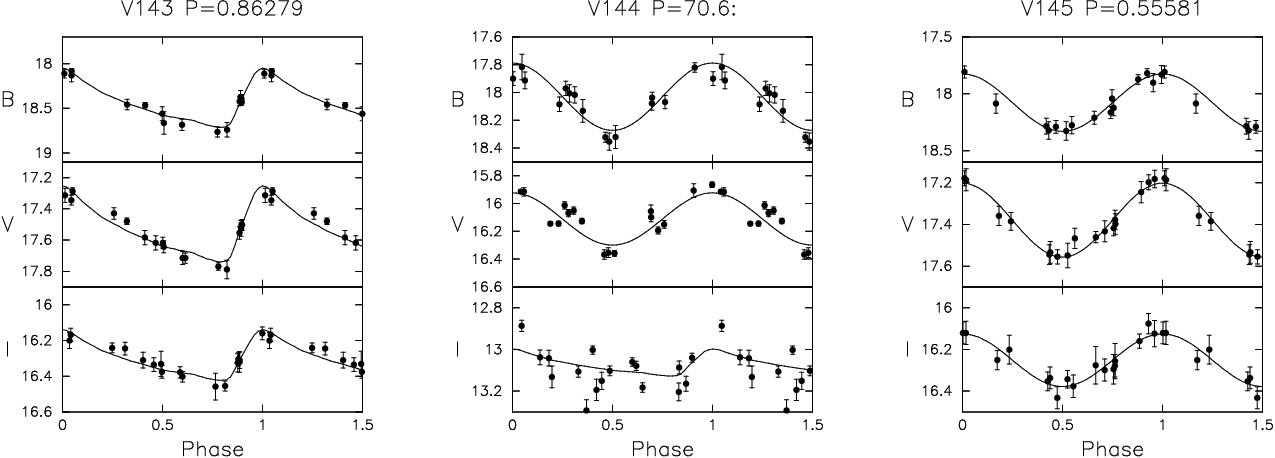,height=2.75in,width=7.0in,angle=-90}} 
  \caption{NGC~6441 light curves in $B$, $V$, and $I$.  Lines shown are 
           template fits to the data.  In some cases, mostly with the 
           $I$ data, the template was not able to find a good fit due 
           to the quality of the data. Long period variables do not have 
           good fits since the templates were not created to deal with 
           this type of variable.} 
  \label{Fig04}
\end{figure*}

We were able to find 38 candidate RRL stars in the inner region
of NGC~6441, with 24 of them being new.  Previous ground-based 
surveys (Layden et al.\ 1999; Pritzl et al.\ 2001) of the cluster found 
another 25 RRL stars which are not included in the field coverage of the 
present study.  We also confirmed the variability of three stars whose 
variablility had been suspected, but not confirmed.  Thus, the 
total number of RRL stars known in NGC~6441 with definite classification 
and membership is now 63.  Figure~4 shows the light curves of the 57 
variables we identified where Layden's template-fitting 
routines were used to fit the data.  Tables~2, 3, and 4 list the $B,V,I$ 
photometric data for the variable stars.  The mean properties of the 
variable stars are given in Table~5.  The mean periods for the 26 fundamental 
(RRab) and 12 RRc stars found in this survey are 0.753~d and 0.365~d, 
respectively.  These values match up very well with those found from the 
ground-based surveys, which were $\langle P_{ab} \rangle = 0.75$~d and 
$\langle P_c \rangle = 0.38$~d for 24 and 9 stars, respectively (Pritzl et 
al.\ 2001).  The variables listed in Table~5 from V55 to V102, were included 
in both the ground-based and HST surveys, so that the two samples are not 
completely independent.  Combining the periods of the 
RRL stars with good light curves, accurate classifications, and 
certain membership found in this survey and the ground-based survey, 
we find the average periods for 42 RRab and 21 RRc stars in NGC~6441 
to be 0.759~d and 0.375~d, respectively.  The ratio of RRc to RRL stars 
is 0.33.  As noted in Pritzl et al.\ (2000, 2001), the mean period of 
the RRab stars is higher than that found in the typical Oosterhoff~II 
cluster.  The mean period of the RRc stars is similar to that for 
Oosterhoff~II clusters, while the $N_c/N_{\rm RR}$ ratio is somewhere between 
the values found for Oosterhoff~I and II clusters.  

Considering only the RRab stars newly discovered in this survey, we 
find a mean period of 0.765~d.  This is slightly larger than one might 
expect given the result from the ground-based 
surveys (Layden et al.\ 1999; Pritzl et al.\ 2001).  Since the HST 
observations allowed us to better survey the 
inner region of NGC~6441, it may be that there is a difference between 
the mean properties of the RRL in the inner and outer regions.  To 
make a more quantitative argument, we find the RRab stars within 20 
arcsec of the cluster center to have a mean period of 0.766~d (17 RRab 
stars).  Those outside this assumed radius have a mean period of 
0.754~d (25 stars).  The radius of 20~arcsec was chosen as the best 
to illustrate the difference in mean period.  For radii less than this 
we begin to enter small-number statistics and for more than this 
the mean periods progressively get closer to matching.

\begin{figure*}[t] 
  \figurenum{5}
  \centerline{\psfig{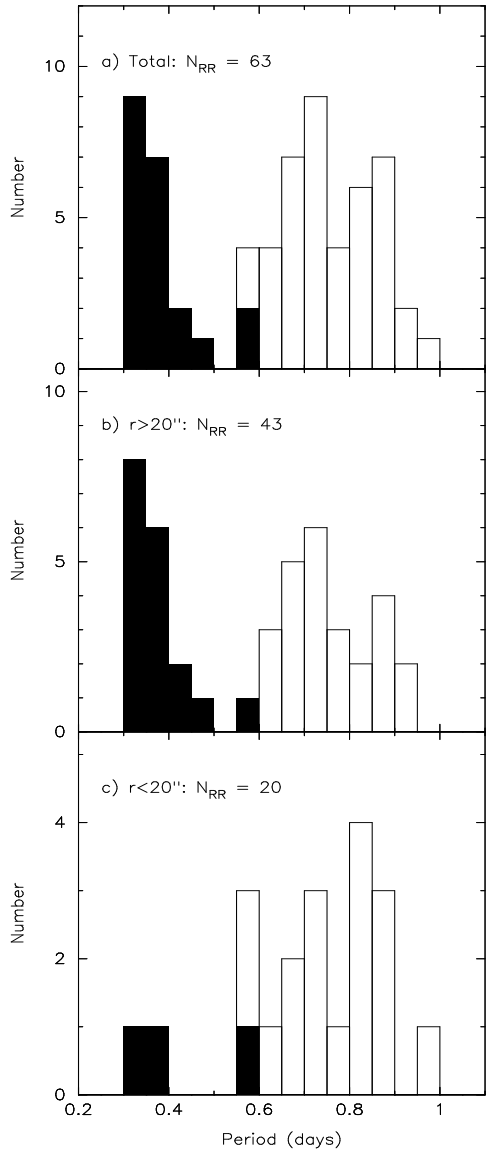}} 
  \caption{NGC~6441 RR~Lyrae histograms for a) all known RR~Lyrae, b) 
           those outside 20~arcsec, and c) those within 20~arcsec.  The 
           filled bars represent the RRc stars, while the unfilled ones 
           are the RRab stars.  In (a) we see that NGC~6441 has the 
           typical Oosterhoff Type~II distribution with a higher proportion 
           of RRc-to-RRab stars as compared to Oosterhoff Type~I 
           clusters whose RR~Lyrae populations are primarily RRab stars.  
           There is a small shift toward longer period for the RR~Lyrae 
           inside 20~arcsec.} 
  \label{Fig05} 
\end{figure*}

Another way to view the shift in period is shown in Figure~5.  The 
histogram in Fig.~5a illustrates that the distribution of RRL stars 
in NGC~6441 is closer to that of an Oosterhoff~II cluster than that of 
an Oosterhoff~I cluster, as discussed in Pritzl et al.\ (2000, 2001).  
A comparison of Fig.~5b and 5c shows an excess of longer period 
RRab stars within 20~arcsec of the cluster center.  Some caution must be 
exercised, however, before one can conclude that the difference is 
real.  A two-sample K-S test run on the inner and outer RRab 
results in a $D$ statistic of 0.2259, which means that the two distributions 
are the same at a 62.1\% confidence level.  
A Student $t$-test, which specifically tests if two distributions are 
different based on their means, was also run and resulted in a $t$ value 
of 0.411, which means that the two distributions are indistinguishable 
at a 68.3\% confidence level.  This confirms the K-S test, thus suggesting 
that the two samples are not significantly different, in a statistical 
sense.  While our HST observations have been able 
to survey the inner regions of NGC~6441, crowding in the regions just 
outside this survey area may have prevented low amplitude, long period 
RRab stars from being detected in the ground-based surveys.  As an 
example, adding five stars with periods of 0.85~d outside 20~arcsec 
would change the mean period to match that found for the RRab inside 
20~arcsec.  We also note that if V118 is truly a P2C (see \S4) and not an 
evolved RRL star, the mean period for the RRab stars within 20~arcsec 
would be 0.753~d.  This matches up very well with 
the mean period for the RRab stars outside 20~arcsec and argues that 
there may be no significant difference between the RRab populations inside 
and outside 20~arcsec.  Further observations of NGC~6441 with better resolution 
at larger radii may help clear up any distinctions 
between the inner and outer RRab stars.  Alternately, application of 
the ISIS image subtraction package (Alard 2000) may improve the 
detection of low amplitude variables in the crowded regions of 
NGC~6441 observed from the ground.  We did not discuss the RRc stars 
given their relatively low numbers.

We derive the mean magnitude for the RRL stars to be $m_{V, {\rm RR}} 
=17.54\pm0.02$~mag, where the error is the standard deviation of the 
mean.  This matches up well with what was found in the ground-based 
survey of Pritzl et al.\ (2001; $m_{V, {\rm RR}}=17.51\pm0.02$~mag).  
As seen in Figure~3, a number of the candidate RRL stars are found at 
magnitudes brighter than the majority of the RRL stars. These variable 
stars have periods of 0.90582~d (V56), 0.76867~d (V110), 0.61419~d 
(V112), 0.97923~d (V118), 0.80573~d (V136), and 0.55581 (V145).  
It may be that differential reddening is playing a role given 
that there is a small number of RRL stars fainter than the majority 
of the RRL stars.  It has already been shown from the RRL star 
light curves that differential reddening affects the ground-based 
observations of NGC~6441 (Layden et al.\ 1999; Pritzl et al.\ 2001).  
We did not attempt to determine the reddenings from the RRab stars in 
this survey via their $\bv$ color at the minimum (Blanco 1992) due to 
the small number of data points.  It may also be the case that some of 
these stars have evolved further from the zero-age HB and are thus more 
luminous.  The periods for two of the stars, V118 (and, to a lesser 
extent V56) and V145, are long for RRab and RRc stars, respectively.  
In fact, given the magnitude and the period of V118, it may be the 
faintest P2C in the cluster (see \S4).

\subsection{Period-Amplitude Diagram}

\begin{figure*}[t]
  \figurenum{6}
  \centerline{\psfig{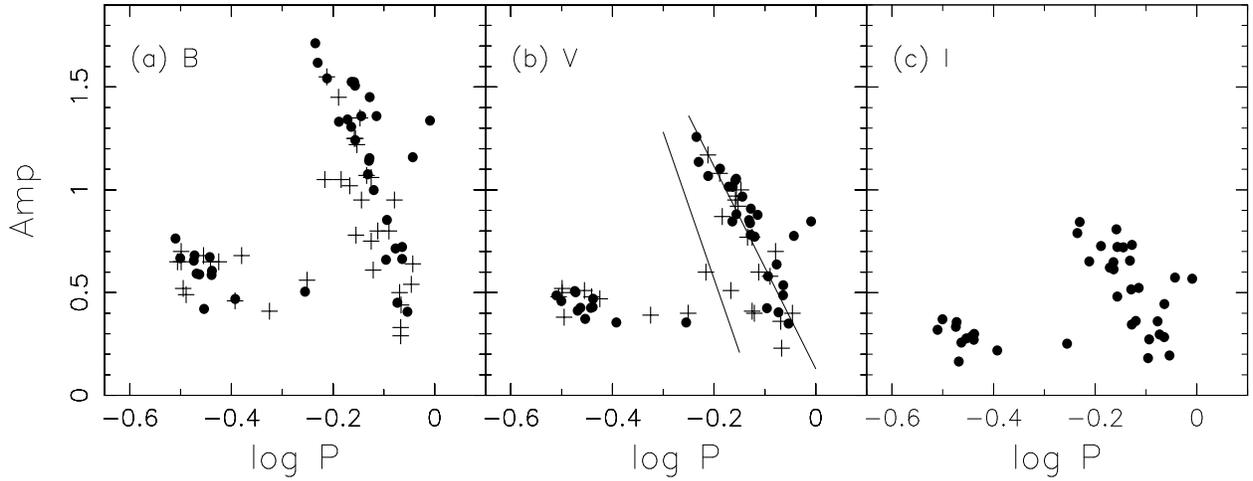}}
  \caption{NGC~6441 period-amplitude diagrams for the RR~Lyrae stars.  
           RR~Lyrae stars found in this survey are shown as filled circles, 
           while those found in the ground-based survey of Pritzl et al.\ 
           (2001) are shown as plusses. There are no clear differences 
           between the RR~Lyrae stars found in the inner and outer regions 
           of NGC~6441.  RRab stars in the ground-based survey scattered 
           toward lower amplitude for their period were likely blended 
           with other stars.  The two lines shown in (b) represent the mean 
           locus for the RRab stars in Oosterhoff~I and Oosterhoff~II 
           globular clusters (Clement 2000).  The NGC~6441 RRab stars fall 
           nearer the Oosterhoff~II line even though according to the 
           metallicity of NGC~6441 it should be an Oosterhoff~I cluster.} 
  \label{Fig06}
\end{figure*}

Figure~6 shows the period-amplitude diagrams for the newly discovered
RRL stars and those previously found (Pritzl et al.\ 2001).  
In general, the RRL stars from the two data sets lie in the same regions
of the diagrams. There is more of a spread among the RRab in
Fig.~6c because the $I$-band light curves have more scatter than the $B$ 
and $V$ light curves.  From Figure~6 we see that the general properties 
of the RRL stars found in the inner region of the cluster are 
the same as those found in the outer regions.  

It was noted in Layden et al.\ (1999) and Pritzl et 
al.\ (2001) that some of the RRab stars were likely blended with other 
stars, producing lower than expected amplitudes for their periods.  
Such blends are likely given the compact nature of this cluster.  The 
candidate blended stars from the ground-based surveys can be seen 
in Figure~6 as those RRab stars which are scattered toward lower 
amplitudes.  In Table~6, we show the comparison between the variable 
stars found in both this survey and the ground-based survey of Pritzl et al.  
There are clearly a number of stars which appeared significantly brighter 
in the ground-based survey and had lower amplitudes than found in the present  
survey.  We also examined the images for stars neighboring those 
listed in Table~6.  All of the stars listed which showed significant 
differences in their magnitudes and periods were found to have nearby 
stars that likely affected the photometry of the ground-based studies.  
The higher resolution of the WFPC2 indicates that the previous conclusion 
of the ``low amplitude" RRab stars in the ground-based survey being 
blends was correct.  There is also the possibility that some of these 
stars may be experiencing the Blazhko effect which also affects the 
amplitudes of RRL stars.  

Layden et al.\ (1999) and Pritzl et al.\ (2000, 2001) found 
that the RRab stars in NGC~6441 lie at longer periods
in the period-amplitude diagram compared to field RRab stars of
similar metallicity. This confirms that the RRL stars, and
therefore the HB, are unusually bright for their metallicity.
Pritzl et al.\ (2001) also showed that the 
RRab stars in NGC~6441 fall near a line in the period-amplitude diagram 
established by Clement (2000) as defining the location of RRab stars 
in Oosterhoff type~II clusters, a result first noted by Clement and 
reconfirmed here in Fig.~6b.  
As noted above, this is consistent with the mean period of the RRab 
actually being more like that of an Oosterhoff type~II cluster than an 
Oosterhoff type~I cluster (see also Walker 2000).  As discussed in detail 
in Pritzl et al.\ (2002), it is difficult to model such GCs as NGC~6388 
and NGC~6441 as Oosterhoff~II systems, under the hypothesis that their 
variables are evolved from a position on the blue zero-age HB, due to the 
small number of progenitors on the blue HB (see also Fig.~3).

We also find one RRc star of unusually long period in the inner region of
NGC~6441 (V145). Two similarly long period
RRc stars were found in the ground-based studies (Layden et al.\ 1999;
Pritzl et al.\ 2001).  In Figure~6 we see that most of the RRc
stars group together, while the longer period ones are seen not to 
partake in the more usual, inverted parabola-like (e.g., Bono et al.\ 
1997a) locus of RRc stars in the period-amplitude plane, for reasons 
which are still unclear.  As noted for NGC~6441 in
Pritzl et al.\ (2001), and for the long period RRc stars in NGC~6388 in
Pritzl et al.\ (2002), there are very few GCs that produce this
type of long period RRc star.

In Figure~6 we can also see that there are two candidate RRab
stars with longer periods than expected for their amplitudes 
(V56 and V118). As noted above, these are two of the stars
which are brighter than the majority of the other RRL stars.
This prompts the question as to whether these stars are
evolved RRL stars or possibly P2Cs. We will discuss the issue
further in the following section.

\subsection{$I$-band Period-Luminosity Relations}

\begin{figure*}[t]
  \figurenum{7}
  \centerline{\psfig{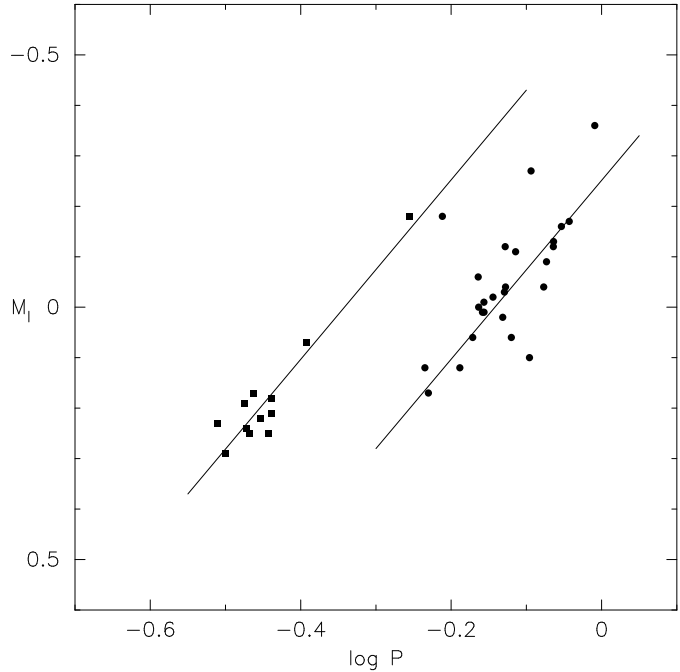}}
  \caption{$I$-band period-luminosity plot for the RR~Lyrae stars in 
           NGC~6441.  The lines shown were derived after excluding the 
           outlying RR~Lyrae stars.} 
  \label{Fig07} 
\end{figure*}

RRL stars provide excellent distance indicators given their 
nearly uniform absolute magnitudes.  However, while the HB 
is flat at the RRL level in $B$ and $V$, thus destroying 
any significant period-luminosity (P-L) relation that might 
otherwise be present due to the important range in $T_{\rm eff}$ 
compared to $L$, the same may not apply in redder (and bluer) 
bandpasses, where the HB becomes significantly non-horizontal 
at the RRL level.  Many studies have been done to determine 
the $K$-band P-L relations for RRL stars (Longmore et al.\ 1990; 
Nemec, Nemec, \& Lutz 1994; Bono et al.\ 2001).  The RRL P-L 
relations in the $I$-band have, on the other hand, been neglected.  
We searched through the literature and found only Layden \& 
Sarajedini (2003) derived a P-L relation in the I-band for the 
RRL in NGC~3201.  Here we 
derive the $I$-band P-L relations for the RRL in NGC~6441 using the 
large number of RRL stars we found in this cluster.

As discussed here, in Layden et al.\ (1999), and in Pritzl et 
al.\ (2001, 2002), the RRL stars in NGC~6388 and NGC~6441 are 
unusually bright for their metallicity.  Given that the general 
properties of the RRL stars in both of these clusters are more 
like Oosterhoff~II clusters than Oosterhoff~I, we assume that 
their RRL stars have absolute magnitudes similar to those of 
Oosterhoff~II clusters such as M15.  Using Eqn.~7 in Lee, Demarque, 
\& Zinn (1990) and adopting a metallicity of ${\rm [Fe/H]}=-2.0$, 
we find an absolute magnitude for the RRL stars in NGC~6441 to be 
$M_V=+0.48$~mag.  Our adopted magnitude for RRL stars in Oosterhoff 
Type~II clusters is broadly consistent with the recent parallax 
determination for RRL stars by Benedict et al.\ (2002). They 
obtained $M_V = +0.61$ for RR~Lyrae itself, which, at 
${\rm [Fe/H]} = -1.4$, is more metal-rich and might be slightly 
fainter than RRL stars in a more metal-poor system.  In 
calculating the distance modulus for NGC~6441, we adopt 
$m_{V,{\rm RR}}=17.54$~mag, $E(\bv)=0.51$, and 
$R_V=3.1$ to find $\mu_0=15.48$~mag.  To derive the individual 
absolute magnitudes we used $A_I=1.85E(\bv)$.  Figure~7 shows the 
P-L relations for the RRL stars.  Removing outlying stars, we 
determine the P-L relations to be,

\begin{equation}
M_{I,{\rm ab}}=-0.25(\pm0.04)-1.78(\pm0.12)\log\,P   
\end{equation}
\begin{equation}
M_{I,{\rm c}}=-0.61(\pm0.06)-1.78(\pm0.14)\log\,P, 
\end{equation}

\noindent
with correlation coefficients $r=0.792$ and 0.947, respectively.  The 
root-mean-square deviations for the RRab and RRc relations are 0.06 and 
0.04, respectively.  As noted in \S3, it is uncertain whether the outlying 
stars, which arebrighter than the majority of RRL stars of similar period, 
are evolved or not.  Still, it is clear that the $I$-band provides a good 
filter to derive P-L relations for the RRL stars.  We note that the zero-point 
for the above relations is uncertain due to the fact that we have assumed 
that the RRL stars in NGC~6441 have the same luminosity as Oosterhoff~II 
RRL stars with ${\rm [Fe/H]}=-2.0$.  The precise luminosity of the NGC~6441 
RRL stars is not conclusively known given the unusual nature of its HB.

\section{Population~II Cepheids}

In the ground-based photometry we were able to find four candidate P2Cs in
NGC~6388 (Pritzl et al.\ 2002), but none in NGC~6441 (Layden
et al.\ 1999; Pritzl et al.\ 2001). At the time it seemed
strange that in two clusters with such similar HB morphology
and RRL star populations, one would contain P2Cs and the other would
not. In this HST snapshot survey we have found six definite P2C
candidates in NGC~6441. This makes NGC~6441, along with NGC~6388,
the most metal-rich GC known to contain P2Cs. Interestingly, 
the P2Cs in both clusters have been found near the inner regions of the 
clusters, though this may reflect just the comparatively small population 
of P2C stars compared to RRL variables.  The discovery of P2Cs in these 
two clusters is consistent with the idea that these types of variables 
are found only in clusters with blue HB components (Wallerstein 1970; 
Smith \& Wehlau 1985). \footnote{Palomar~3 is a notable exception 
to this otherwise seemingly general rule (Borissova, Ivanov, \& 
Catelan 2000).}

\begin{figure*}[t]
  \figurenum{8}
  \centerline{\psfig{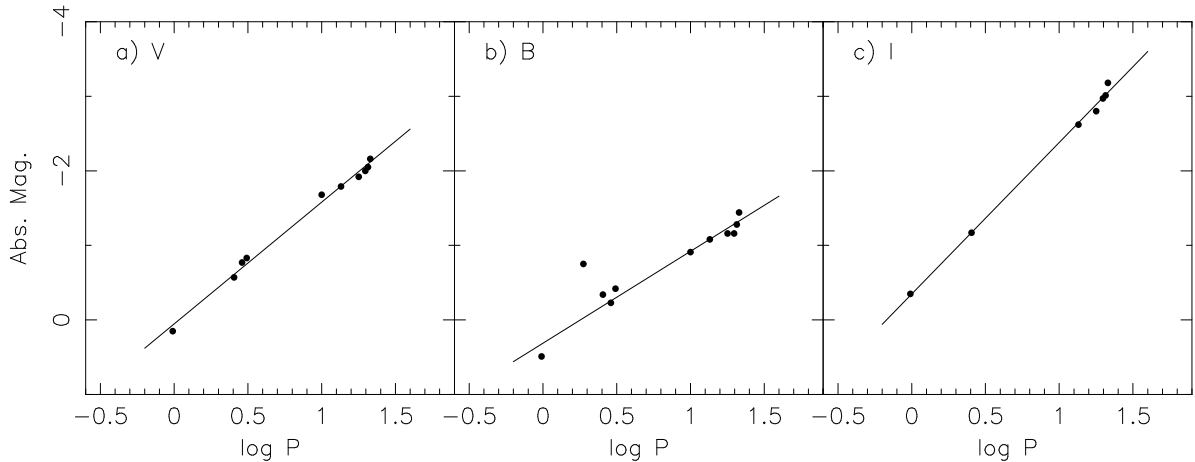}}
  \caption{Period-luminosity plots for the Population~II Cepheids in 
           NGC~6441 and NGC~6388. All of the relations are shown to be 
           exceptionally tight, with the exception of one star in $B$ 
           which is thought to be saturated (see text and Pritzl et al.\ 
           2002).  The correlation coefficients are 0.985, 0.996, and 
           0.998 for $B$, $V$, and $I$, respectively.} 
  \label{Fig08}
\end{figure*}

Of the six P2Cs, five are of W~Virginis type (W~Vir; $10 < P < 20$~d)
and one is of BL~Herculis type (BL~Her; $P<10$~d). The difference
between these types is not only defined by their periods, but also by
their origins. As reviewed by Wallerstein (2002, but see also 
Bono et al.\ 1997b), BL~Her stars are 
believed to derive from stars which have evolved from the HB bluer than 
the RRL gap, and which are now evolving redward through the 
instability strip at luminosities brighter than those of RRL stars.
By contrast, W~Vir stars, while also considered to be the progeny 
of blue HB stars (e.g., Gingold 1985), are believed to be undergoing helium shell flashes
leading them along blueward loops into the instability strip as they move up 
the asymptotic giant branch (AGB).\@  This is consistent with studies of period
changes in P2Cs in GCs. Wehlau \& Bohlender (1982) found that 
of 12 BL~Her stars in Galactic GCs, nine had increasing periods, three had unchanging 
periods, and zero had decreasing periods. Increasing periods are expected 
for variables evolving redward from the blue HB toward the AGB.\@  For the 
W~Vir stars the picture is not so clear, but Clement, Hogg, \& Yee (1988) 
found that a number of W~Vir stars showed both increases and decreases 
in period, and that at least two variables showed long term period 
decreases consistent with blueward evolution. The period change 
observations for W~Vir stars are thus at least consistent with evolution 
of W~Vir stars into and out of the instability strip during blueward loops.

\subsection{Period-Luminosity Relation for Population II Cepheids}

Given the relatively large number of P2Cs found in both NGC~6441 and
NGC~6388, we would like to determine the P-L relations 
for these stars.  We use the distance modulus derived in \S3.2 for 
NGC~6441, $\mu_0=15.48$~mag, to derive the absolute magnitudes for 
the P2Cs found in this survey.  For NGC~6388, we follow the prescription 
used for NGC~6441 to derive its distance modulus.  Since we assume 
that the RRL stars in NGC~6388 are of a similar nature as those in 
NGC~6441, we adopt the same absolute magnitude for its RRL stars, 
$M_V=+0.48$~mag. 

In Table~7 we list the data for the P2Cs in NGC~6388 and NGC~6441.  For
our discussions here, we will include V118 with the P2Cs, although its
precise classification is uncertain.  We have chosen not to include V56
since its period and magnitude are not unreasonably different
from the other RRL stars.  We have assumed
$R_V=3.1$, $R_B=4.1$, and $A_I=1.85 E(\bv)$, and have adopted
mean reddenings of $E(\bv)=0.40\pm0.03$ and $0.51\pm0.02$ for NGC~6388 and
NGC~6441, respectively (Pritzl et al.\ 2001, 2002).  Figure~8 shows
the absolute magnitudes of the P2Cs in NGC~6388 and NGC~6441 for the various
filters. In each panel of Figure~8, we see that the P2Cs form a surprisingly 
good correlation from shorter to longer periods, with the lone exception 
of one star in NGC~6388.  V29 of NGC~6388 was shown to be unusually bright for
its period in Pritzl et al.\ (2002) leading to the idea that it may
be blended with another star.  Leaving that star out, we determined the 
following P-L relations, 

\begin{equation}
M_V=0.05(\pm0.05) - 1.64(\pm0.05) \log\,P  
\end{equation}
\begin{equation}
M_B=0.31(\pm0.09) - 1.23(\pm0.09) \log\,P 
\end{equation}
\begin{equation}
M_I=-0.36(\pm0.01) - 2.03(\pm0.03) \log\,P, 
\end{equation}

\noindent
with correlation coefficients $r=0.996$, 0.985, and 0.998, respectively.  The 
root-mean-square deviation for each relation is 0.07, 0.10, 
and 0.06 for $V$, $B$, and $I$, respectively.  
Both the slopes for $M_V$ and $M_B$ match up quite well with those
found by McNamara (1995, Eqns.~11 and 12) for P2Cs with periods less
than ten days, but are very different compared to the results of
Alcock et al.\ (1998, Eqn.~5) for longer period P2Cs.  The tight fits
to the relations also indicate that all of the P2Cs in NGC~6388 and
NGC~6441 are likely to be pulsating in one mode, presumably the
fundamental mode.  Following the suggestion by Alcock et al.\ that 
adding a color term results in better relations, we calculated the 
P-L-C relations for the NGC~6441 and NGC~6388 P2Cs and found no 
significant improvement in the quality of the fit.  This may 
be explained by the already excellent correlation of the P2Cs in the P-L 
relations and the fact that Alcock et al.\ found the RV~Tauri (RV~Tau; 
$P>20$~days) benefited the greatest in adding in the color 
term.

\subsection{Comparison of Period-Luminosity Relations} 

Defining a clear P-L relation is important in the use of
Cepheid variable stars in distance determinations.  While other Cepheid-type
variables such as the classical and anomalous types have their P-L relations
well-defined, the relations for the P2Cs in general are still somewhat
uncertain.  Arp (1955a) argued that P2Cs follow at least two parallel
P-L relations, one each for fundamental and overtone mode pulsators. Others 
later favored a single P-L relation to describe the P2Cs (e.g., Dickens \& 
Carey 1967; Demers \& Harris 1974). Harris (1985) also assumed a single P-L
relation, but with a break toward a steeper slope for P2Cs with periods
greater than 12.5~days.  The theoretical $B$-band P-L relations of 
Bono et al.\ (1997b) show a break in slope near $P=15$~days at the 
faint boundary of the domain in which P2Cs might exist, but no break 
in slope at the brighter boundary.

\begin{figure*}[t]
  \figurenum{9}
  \centerline{\psfig{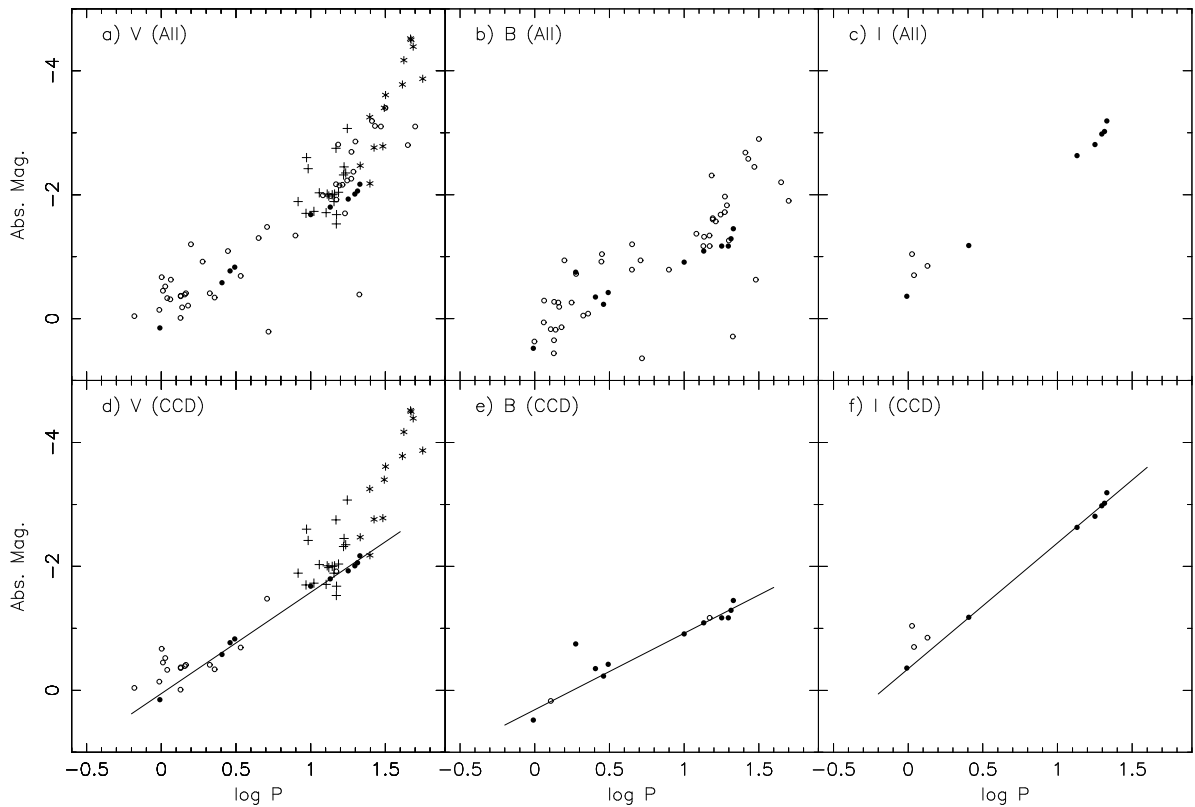}} 
  \caption{Period-luminosity plots for Population~II Cepheids. In the top 
           row (a, b, c) we include all the data available for Population~II 
           Cepheids in Galactic GCs (NGC~6388 and NGC~6441: filled circles; 
           all others: open circles) and the Large Magellanic Cloud 
           (W~Virginis stars: plusses; RV~Tauri: asterisks).  The bottom 
           row (d, e, f) shows only those stars with CCD photometry.  We 
           also include in these figures the period-luminosity lines found 
           for the NGC~6388 and NGC~6441 Population~II Cepheids.} 
  \label{Fig09}
\end{figure*}

In their paper on Population~II variable stars, Nemec, Nemec \& Lutz (1994)
revived Arp's model stating that P2Cs are pulsating in either fundamental
or first overtone modes, which followed their theme that all pulsating
variable stars can be found to have both pulsation modes.  However, McNamara
(1995), using the data from Nemec et al., stated that the 
observations were better fit by one line describing the BL~Her stars 
($P<10$~d) and another fitted to the W~Vir stars ($P>10$~d), similar to 
what was argued by Harris (1985). Recently, Alcock et al.\ (1998) used 
a sample of 33 P2Cs in the Large Magellanic Cloud (LMC) in the period 
range of $0.9 < \log\,P < 1.75$ to compare the various P-L relations.  
While noting that the Galactic GC P-L relations provided adequate fits to 
the shorter period LMC P2Cs, they found that those P-L 
relations did not provide a good fit for the longer period RV~Tau stars, 
which appear more luminous than the P-L relations predicted.  Combining the
Galactic GC and LMC data sets, Alcock et al.\ found that the shorter
period Galactic GC P2Cs provided a good linear extension to the LMC data with 
the RV~Tau stars turning toward brighter magnitudes.  To account for this,
Alcock et al.\ included a color term to their P-L relations for $\log\,P > 
0.9$, which greatly reduced the scatter in the data and the up-turn of 
the RV~Tau stars.  They concluded that the RV~Tau stars are direct 
extensions of the other P2C types.  Unfortunately, the Alcock et al.\ data 
were taken with $V$ and $R$ filters, while most of the Galactic GC data is in 
$B$ and $V$.  This makes the application of their P-L-C relation 
difficult to apply to other data sets.

To help clarify the issue of the differing slopes found by various studies, 
we investigated the P2Cs in other GCs and incorporated them in our 
analysis.  Table~8 lists the properties of the Galactic GC P2Cs summarizing 
the data found in the literature.  We estimated the absolute
magnitudes for each star using Eqn.~7 of Lee et al.\ (1990)
and the metallicities, reddenings, and $V_{\rm HB}$ given in Harris
(1996).  The format is the same as in Table~7.  We plot the best available 
data for the Galactic GC P2Cs in the top row of Figure~9, including 
those found by Alcock et al.\ (1998) in the LMC where they adopted a 
distance modulus of 18.5~mag.  The only P2Cs not included in the plot were 
those from NGC~4372, which are clearly too faint.  The periods 
listed for these stars may be correct, but the faint magnitudes may be 
due to the fact that Kaluzny \& Krzeminski (1993) were searching for 
contact binaries and SX~Phoenicis stars resulting in the candidate P2Cs 
being saturated.  There is some scatter among the P2Cs
in Fig.~9a, which is to be expected since it is known that there is some
width to the instability strip. Fig.~9b shows a lot of scatter which
may possibly be attributed to the presence of older photographic data, as
compared to the $I$ data in Fig.~9c which is only from CCD data. 
We plot in the bottom row of Figure~9 only those P2Cs with CCD photometry, 
except again for NGC~4372.  This helps to reduce the scatter, as seen in 
Fig.~9d, but some still remains.  Without any color information we 
are not able to tell if this scatter may be intrinsic,
possibly due to the width of the P2C instability strip, or if it is 
photometric.  It should be noted that the scatter seen at shorter 
periods in Fig.~9d can be attributed to the $\omega$~Cen P2Cs, a 
number of which have poor light curves (Kaluzny et al.\ 1997).  For 
$B$ and $I$, the CCD data are mostly taken from this study, but it is 
encouraging that the other data are consistent with what we have found.

Taking a closer look at the $B$ and $I$ data, it does not look like there
is a change in the slope between the BL~Her and W~Vir stars. However, this 
is a tentative conclusion given the fact that
most of the data comes from this survey alone. If we examine the $V$ data in
Fig.~9d, where more points are available, we can make some stronger
arguments. There is clearly a change in the slope toward brighter
magnitudes at longer periods. The question is whether this change is found 
at or around 10 or 12 days as suggested by Harris (1985) and McNamara (1995) 
or at a longer period. If we consider only the BL~Her and
W~Vir stars, there is statistically no change in the slope between the
two groups of P2Cs. On the other hand, if we consider the RV~Tau
stars, they clearly have a different slope from that of the shorter
period Cepheids. It appears, therefore, that a single P-L relationship
can be used to describe the P2Cs until periods greater than about 20 days
are reached, when variables with RV~Tau characteristics begin to enter 
the mix.  Unfortunately, given the scarcity of data, we are not able to 
test the conclusion of Alcock et al.\ (1998) that adding in the color 
term to the RV~Tau stars brings them more into line with the other P2C types. 

Another possibility is raised by Fig.~9d.  Alcock et al.\ (1998) identified 
a number of W~Vir stars found in the LMC that are brighter than the bulk 
of the W~Vir stars.  These variables appear to constitute a straight line 
in combination with the brighter RV~Tau stars that appears to be parallel 
to a line that could be drawn through the BL~Her and W~Vir stars.  
Alcock et al (1998) classified their RV~Tau stars as those having periods 
longer than 20~days.  It may be that some of their W~Vir stars are actually 
RV~Tau stars, and vice versa.  There appears to be in reality no sharp
dividing line between the W~Vir and RV~Tau stars, with the longest period 
W~Vir variables gradually taking on RV~Tau characteristics (Alcock et al.\ 
1998).  

At this point, it is uncertain how well the relations created from the 
NGC~6441 and NGC~6388 P2Cs apply to other GCs given the small amount of 
CCD photometry.  Two of the more important questions are how good are 
the zeropoints we are using to derive the absolute magnitudes and is 
there any metallicity dependence in the relations which is currently not 
being taken into account in our relations.  Clearly, more CCD photometry 
needs to be taken of P2Cs in order to clarify these issues, especially 
in the $B$ and $I$ filters, before any firm conclusions can be drawn.

\section{Summary and Conclusions}

We have found 57 variable stars in the inner regions of NGC~6441
using the HST snapshot observing program, 41 of which were previously
undiscovered.  Of the RRL stars found in this survey, 26 are
RRab stars and 12 are RRc stars.  There are no clear differences in the
properties of the RRL stars in the inner and outer regions of the
cluster.  This is illustrated in the period-amplitude diagram where 
the RRL stars in this survey are located in the same regions as 
those found in the ground-based surveys (Layden et al.\ 1999; Pritzl et 
al.\ 2001).  Combining the RRL stars in this survey and those found 
in ground-based surveys, we find the mean periods of the RRab and RRc 
stars to be 0.759~d and 0.375~d.  This reaffirms that NGC~6441 is unusual 
when compared to other Galactic GCs in that, whereas it 
is a metal-rich GC, its RRL stars have mean periods that are similar to, 
if not greater than, those of very metal-poor GCs.  This reconfirms the 
conclusions of Pritzl et al.\ (2001) that the RRL stars in NGC~6441 
are unusually bright for their metallicity.  This breaks the general 
metallicity-luminosity relation for the RRL stars and has consequences 
on their use as distance indicators.  A possible trend of the inner RRab stars 
($r<20$~arcsec) having a longer mean period than the remaining RRab stars 
was also detected, but further observations will be needed to make 
definitive conclusions.  We also used the $I$-band photometry to derive 
P-L relations for the RRL stars.

We also found six P2Cs in NGC 6441 making it, along with its ``sister" 
cluster NGC~6388, the most metal-rich GC to contain this 
type of variable star.  Five of the P2Cs are of W~Vir type
and one is a BL~Her star.  Also, V118 may be an additional BL Her 
variable given its bright magnitude and long period compared to the 
RRL stars.  Under the assumption that the RRL stars in NGC~6388 
and NGC~6441 have absolute magnitudes similar to those of metal-poor 
GCs, we have derived the $B$, $V$, and $I$ period-luminosity 
relations for their P2Cs.  These P-L relations do not show the change 
in slope around $P=10$~day which some have suggested and show a 
surprisingly tight correlation.  We also find that only one P-L relation 
is needed to describe the data, i.e., it appears that the P2Cs 
are pulsating in one mode.  An examination of the P2Cs in GCs 
for which $V$ CCD photometry is available indicates that the 
break in the slope of the P2C P-L relation does not come at periods of 
10 or 12 days, but at longer periods, where the P2Cs begin to
take on RV~Tau characteristics.

\acknowledgments

Support for Proposal number SNAP~8251 was provided by NASA through a grant 
from the Space Telescope Science Institute, which is operated by the 
Association of Universities for Research in Astronomy, Incorporated, 
under NASA contract NAS5-26555.

Support for M.C. was provided by Proyecto de Inicio DIPUC 2002-04E 
and Proyecto FONDECYT Regular 1030954.

\begin{deluxetable}{cccc} 
\tablewidth{0pc}
\tablecaption{NGC~6441 HST Observation Log\label{tbl-1}}
\tablehead{
\colhead{HST Dataset Name} & \colhead{Date} & \colhead{Filter} & 
\colhead{Exposure Times} \\
 & & & \colhead{(sec)} 
          }
\startdata \singlespace
\small U2HJ0103T & \small 1994 August 08 & \small F439W & \small \small 50 \\ 
\small U2HJ0104T & \small 1994 August 08 & \small F439W & \small 500 \\
\small U2VO0801T & \small 1995 September 12 & \small F555W & \small 14 \\ 
\small U2VO0802T & \small 1995 September 12 & \small F555W & \small 50 \\ 
\small U2VO0803T & \small 1995 September 12 & \small F439W & \small 50 \\ 
\small U2VO0804T & \small 1995 September 12 & \small F439W & \small 160 \\ 
\small U2VO0805T & \small 1995 September 12 & \small F439W & \small 160 \\ 
\small U39U010ET & \small 1996 August 8 & \small F547M & \small 80 \\ 
\small U39U010FT & \small 1996 August 8 & \small F547M & \small 80 \\ 
\small U39U010GT & \small 1996 August 8 & \small F547M & \small 20 \\ 
\small U39U010HT & \small 1996 August 8 & \small F547M & \small 20 \\ 
\small U59V1101R & \small 1999 March 16 & \small F439W & \small 50 \\ 
\small U59V1102R & \small 1999 March 16 & \small F555W & \small 10 \\ 
\small U59V1103R & \small 1999 March 16 & \small F814W & \small 10 \\ 
\small U59V1201R & \small 1999 April 6 & \small F439W & \small 50 \\ 
\small U59V1202R & \small 1999 April 6 & \small F555W & \small 10 \\ 
\small U59V1203R & \small 1999 April 6 & \small F814W & \small 10 \\ 
\small U59V0501R & \small 1999 April 18 & \small F439W & \small 50 \\ 
\small U59V0502R & \small 1999 April 18 & \small F555W & \small 10 \\ 
\small U59V0503R & \small 1999 April 18 & \small F814W & \small 10 \\ 
\small U59V0301R & \small 1999 July 15 & \small F439W & \small 80 \\ 
\small U59V0302R & \small 1999 July 15 & \small F555W & \small 10 \\ 
\small U59V0303R & \small 1999 July 15 & \small F814W & \small 10 \\ 
\small U59V0101R & \small 1999 July 25 & \small F439W & \small 80 \\ 
\small U59V0102R & \small 1999 July 25 & \small F555W & \small 10 \\ 
\small U59V0103R & \small 1999 July 25 & \small F814W & \small 10 \\ 
\small U59V2001R & \small 2000 August 19 & \small F439W & \small 80 \\ 
\small U59V2002R & \small 2000 August 19 & \small F555W & \small 10 \\ 
\small U59V2003R & \small 2000 August 19 & \small F814W & \small 10 \\ 
\small U59V0701R & \small 1999 August 21 & \small F439W & \small 80 \\ 
\small U59V0702R & \small 1999 August 21 & \small F555W & \small 10 \\ 
\small U59V0703R & \small 1999 August 21 & \small F814W & \small 10 \\ 
\small U59V0901R & \small 1999 August 23 & \small F439W & \small 80 \\ 
\small U59V0902R & \small 1999 August 23 & \small F555W & \small 10 \\ 
\small U59V0903R & \small 1999 August 23 & \small F814W & \small 10 \\ 
\small U59V0401R & \small 1999 September 2 & \small F439W & \small 80 \\ 
\small U59V0402R & \small 1999 September 2 & \small F555W & \small 10 \\ 
\small U59V0403R & \small 1999 September 2 & \small F814W & \small 10 \\ 
\small U59V1001R & \small 1999 September 4 & \small F439W & \small 80 \\ 
\small U59V1002R & \small 1999 September 4 & \small F555W & \small 10 \\ 
\small U59V1003R & \small 1999 September 4 & \small F814W & \small 10 \\ 
\small U59V1601R & \small 1999 September 16 & \small F439W & \small 80 \\ 
\small U59V1602R & \small 1999 September 16 & \small F555W & \small 10 \\ 
\small U59V1603R & \small 1999 September 16 & \small F814W & \small 10 \\ 
\small U59V0601R & \small 1999 September 19 & \small F439W & \small 80 \\ 
\small U59V0602R & \small 1999 September 19 & \small F555W & \small 10 \\ 
\small U59V0603R & \small 1999 September 19 & \small F814W & \small 10 \\ 
\small U59V1901R & \small 1999 September 19 & \small F439W & \small 80 \\ 
\small U59V1902R & \small 1999 September 19 & \small F555W & \small 10 \\ 
\small U59V1903R & \small 1999 September 19 & \small F814W & \small 10 \\ 
\small U59V0201R & \small 2000 February 29 & \small F439W & \small 80 \\ 
\small U59V0202R & \small 2000 February 29 & \small F555W & \small 10 \\ 
\small U59V0203R & \small 2000 February 29 & \small F814W & \small 10 \\ 
\small U59V1401R & \small 2000 March 4 & \small F439W & \small 80 \\ 
\small U59V1402R & \small 2000 March 4 & \small F555W & \small 10 \\ 
\small U59V1403R & \small 2000 March 4 & \small F814W & \small 10 \\ 
\small U59V1701R & \small 2000 May 23 & \small F439W & \small 80 \\ 
\small U59V1702R & \small 2000 May 23 & \small F555W & \small 10 \\ 
\small U59V1703R & \small 2000 May 23 & \small F814W & \small 10 \\ 
\small U59V1301R & \small 2000 May 26 & \small F439W & \small 80 \\ 
\small U59V1302R & \small 2000 May 26 & \small F555W & \small 10 \\ 
\small U59V1303R & \small 2000 May 26 & \small F814W & \small 10 \\ 
\small U59V0801R & \small 2000 June 30 & \small F439W & \small 80 \\ 
\small U59V0802R & \small 2000 June 30 & \small F555W & \small 10 \\ 
\small U59V0803R & \small 2000 June 30 & \small F814W & \small 10 \\ 
\enddata
\end{deluxetable}

\begin{deluxetable}{cccc}
\tablewidth{0pc}
\tablecaption{NGC~6441 $B$ Photometry of Variable Stars \label{tbl-2}} 
\tablehead{
\colhead{ID} & \colhead{HJD} & \colhead{$B$} & \colhead{$\sigma_B$} 
          }
\startdata
V001 & 51374.930 &  20.830  &  0.078 \\
V001 & 51410.434 &  20.476  &  0.145 \\
V001 & 51411.773 &  20.376  &  0.111 \\
V001 & 51413.652 &  20.225  &  0.102 \\
V001 & 51424.422 &  19.308  &  0.075 \\
V001 & 51425.836 &  19.194  &  0.069 \\
V001 & 51437.648 &  18.445  &  0.067 \\
V001 & 51440.871 &  18.418  &  0.100 \\ 
V001 & 51440.949 &  18.347  &  0.044 \\
V006 & 49972.754 &  16.118  &  0.052 \\
V006 & 49972.754 &  16.118  &  0.052 \\
V006 & 49972.754 &  16.123  &  0.046 \\
V006 & 49972.754 &  16.123  &  0.046 \\
V006 & 49972.758 &  16.120  &  0.041 \\
V006 & 51374.930 &  16.974  &  0.108 \\
V006 & 51385.414 &  15.449  &  0.139 \\
V006 & 51410.434 &  15.640  &  0.038 \\
V006 & 51411.773 &  15.791  &  0.025 \\
V006 & 51413.652 &  16.036  &  0.041 \\
V006 & 51424.422 &  16.173  &  0.130 \\
V006 & 51425.836 &  15.774  &  0.112 \\
V006 & 51440.871 &  17.316  &  0.042 \\
V006 & 51440.949 &  17.340  &  0.099 \\
\enddata
\tablecomments{The complete version of this table is in the electronic 
edition of the Journal. The printed edition contains only a sample.}
\end{deluxetable}

\begin{deluxetable}{cccc}
\tablewidth{0pc}
\tablecaption{NGC~6441 $V$ Photometry of Variable Stars \label{tbl-3}} 
\tablehead{
\colhead{ID} & \colhead{HJD} & \colhead{$V$} & \colhead{$\sigma_V$} 
          }
\startdata
V001 & 51374.934  & 18.945  &  0.072 \\
V001 & 51385.418  & 19.091  &  0.059 \\
V001 & 51411.777  & 18.589  &  0.092 \\
V001 & 51413.656  & 18.451  &  0.095 \\
V001 & 51424.422  & 17.588  &  0.046 \\
V001 & 51425.836  & 17.471  &  0.050 \\
V001 & 51437.648  & 16.847  &  0.023 \\
V001 & 51440.871  & 16.765  &  0.079 \\
V001 & 51440.949  & 16.702  &  0.084 \\
V006 & 49972.746  & 14.681  &  0.056 \\
V006 & 49972.750  & 14.744  &  0.084 \\
V006 & 51374.934  & 15.470  &  0.049 \\
V006 & 51385.418  & 14.311  &  0.086 \\
V006 & 51410.434  & 14.701  &  0.310 \\
V006 & 51411.777  & 14.659  &  0.073 \\
V006 & 51413.656  & 14.864  &  0.065 \\
V006 & 51424.422  & 15.061  &  0.080 \\
V006 & 51425.836  & 14.817  &  0.067 \\
V006 & 51440.871  & 15.557  &  0.043 \\
V006 & 51440.949  & 15.580  &  0.042 \\
V006 & 51725.844  & 14.497  &  0.060 \\
\enddata
\tablecomments{The complete version of this table is in the electronic 
edition of the Journal. The printed edition contains only a sample.}
\end{deluxetable}

\begin{deluxetable}{cccc}
\tablewidth{0pc}
\tablecaption{NGC~6441 $I$ Photometry of Variable Stars \label{tbl-4}} 
\tablehead{
\colhead{ID} & \colhead{HJD} & \colhead{$I$} & \colhead{$\sigma_I$} 
          }
\startdata
V001 & 51374.934  & 13.340  &  0.045 \\
V001 & 51385.422  & 13.503  &  0.042 \\
V001 & 51410.438  & 13.366  &  0.064 \\
V001 & 51411.777  & 13.248  &  0.077 \\
V001 & 51413.656  & 13.224  &  0.074 \\
V001 & 51424.426  & 12.800  &  0.056 \\
V001 & 51425.840  & 12.761  &  0.057 \\
V001 & 51437.652  & 12.918  &  0.108 \\
V001 & 51440.875  & 12.687  &  0.091 \\
V001 & 51440.953  & 12.654  &  0.074 \\
V006 & 51374.934  & 13.786  &  0.029 \\
V006 & 51385.422  & 12.876  &  0.065 \\
V006 & 51410.438  & 12.863  &  0.058 \\
V006 & 51411.777  & 12.897  &  0.045 \\
V006 & 51413.656  & 13.081  &  0.035 \\
V006 & 51424.426  & 13.577  &  0.041 \\
V006 & 51425.840  & 13.375  &  0.066 \\
V006 & 51440.875  & 13.761  &  0.019 \\
V006 & 51440.953  & 13.758  &  0.031 \\
\enddata
\tablecomments{The complete version of this table is in the electronic 
edition of the Journal. The printed edition contains only a sample.}
\end{deluxetable}

\begin{deluxetable}{ccccccccccccl}
\tablewidth{0pc}
\tabletypesize{\footnotesize}
\tablecaption{NGC~6441 Variable Star Mean Properties \label{tbl-5}}
\tablehead{
\colhead{ID} & \colhead{RA} & \colhead{Dec} & \colhead{Period (d)} & 
\colhead{$\langle B \rangle$} & 
\colhead{$A_B$} & \colhead{$\langle V \rangle$} & \colhead{$A_V$} & 
\colhead{$\langle I \rangle$} & \colhead{$A_I$} & 
\colhead{Classification} 
          }
\startdata 
001 & 17:50:17.13 & -37:03:50.3 & 89.9:    & 19.853 & 2.691 & 17.848 & 2.633 & 13.092 & 0.924 & LPV \\
006 & 17:50:15.65 & -37:02:16.3 & 21.365   & 16.117 & 2.017 & 14.885 & 1.274 & 13.231 & 1.125 & P2C \\
010 & 17:50:19.54 & -37:04:05.0 & 155:     & 18.958 & 1.897 & 16.618 & 1.817 & 12.843 & 1.079 & LPV \\
017 & 17:50:07.36 & -37:02:40.7 & 257.6    & 18.712 & 0.985 & 16.802 & 0.938 & 13.648 & 0.285 & LPV \\
040 & 17:50:10.54 & -37:04:00.6 & 0.64800  & 18.397 & 1.331 & 17.572 & 1.103 & 16.541 & 0.727 & RRab \\
055 & 17:50:12.46 & -37:02:28.2 & 0.69750  & 18.333 & 1.507 & 17.562 & 1.054 & 16.416 & 0.722 & RRab \\
056 & 17:50:11.13 & -37:02:39.9 & 0.90582  & 18.239 & 1.034 & 17.410 & 0.776 & 16.253 & 0.573 & RRab \\
057 & 17:50:10.44 & -37:02:56.1 & 0.69438  & 18.325 & 1.524 & 17.575 & 1.046 & 16.435 & 0.808 & RRab \\
058 & 17:50:14.72 & -37:03:07.2 & 0.68538  & 18.363 & 1.307 & 17.581 & 0.846 & 16.361 & 0.648 & RRab \\
063 & 17:50:11.34 & -37:02:47.4 & 0.69781  & 18.379 & 1.242 & 17.572 & 0.881 & 16.431 & 0.481 & RRab \\
064 & 17:50:09.64 & -37:03:07.7 & 0.71710  & 18.366 & 1.359 & 17.592 & 0.967 & 16.399 & 0.720 & RRab \\
065 & 17:50:13.20 & -37:02:30.2 & 0.75850  & 18.528 & 0.999 & 17.688 & 0.772 & 16.479 & 0.362 & RRab \\
075 & 17:50:12.50 & -37:03:40.6 & 0.40502  & 18.139 & 0.469 & 17.471 & 0.355 & 16.498 & 0.219 & RRc \\
093 & 17:50:06.34 & -37:02:45.1 & 0.34004  & 18.020 & 0.592 & 17.474 & 0.412 & 16.670 & 0.165 & RRc \\
095 & 17:50:08.51 & -37:04:13.8 & 0.089928 & 18.369 & 0.793 & 17.605 & 0.559 & 16.604 & 0.356 & $\delta$~Scuti \\
102 & 17:50:09.63 & -37:04:11.1 & 0.30889  & 18.139 & 0.763 & 17.526 & 0.486 & 16.657 & 0.319 & RRc \\
105 & 17:50:12.32 & -37:02:52.6 & 111.6    & 18.137 & 1.266 & 16.151 & 1.487 & 13.039 & 0.867 & LPV; SV7 \\
106 & 17:50:10.61 & -37:03:26.0 & 0.36092  & 18.304 & 0.673 & 17.650 & 0.426 & 16.670 & 0.287 & RRc; SV14 \\
107 & 17:50:15.24 & -37:03:07.1 & 0.73891  & 18.499 & 1.076 & 17.650 & 0.853 & 16.443 & 0.655 & RRab \\
108 & 17:50:14.45 & -37:02:42.9 & 0.34419  & 18.188 & 0.589 & 17.531 & 0.426 & 16.596 & 0.257 & RRc \\
109 & 17:50:14.72 & -37:02:50.0 & 0.36455  & 18.285 & 0.606 & 17.611 & 0.470 & 16.633 & 0.299 & RRc \\
110 & 17:50:14.07 & -37:03:00.8 & 0.76867  & 18.198 & 1.359 & 17.399 & 0.878 & 16.317 & 0.523 & RRab \\
111 & 17:50:13.69 & -37:02:57.8 & 0.74464  & 18.386 & 1.154 & 17.552 & 0.781 & 16.305 & 0.345 & RRab \\
112 & 17:50:13.61 & -37:02:54.3 & 0.61419  & 18.177 & 1.542 & 17.363 & 1.067 & 16.245 & 0.652 & RRab \\
113 & 17:50:13.56 & -37:02:56.6 & 0.58845  & 18.335 & 1.618 & 17.605 & 1.136 & 16.589 & 0.844 & RRab \\
114 & 17:50:13.46 & -37:02:53.3 & 0.67389  & 18.366 & 1.342 & 17.610 & 1.015 & 16.485 & 0.622 & RRab \\
115 & 17:50:13.27 & -37:02:46.4 & 0.86311  & 18.458 & 0.722 & 17.552 & 0.536 & 16.291 & 0.444 & RRab \\
116 & 17:50:13.12 & -37:03:22.7 & 0.58229  & 18.264 & 1.713 & 17.612 & 1.257 & 16.544 & 0.790 & RRab \\
117 & 17:50:13.10 & -37:03:12.2 & 0.74529  & 18.314 & 1.451 & 17.557 & 0.908 & 16.456 & 0.576 & RRab \\
118 & 17:50:12.50 & -37:03:20.8 & 0.97923  & 18.048 & 1.337 & 17.201 & 0.846 & 16.061 & 0.568 & RRab?;P2C? \\
119 & 17:50:12.45 & -37:03:01.3 & 0.68628  & 18.203 & 1.525 & 17.511 & 1.014 & 16.422 & 0.613 & RRab \\
120 & 17:50:12.19 & -37:02:53.5 & 0.36396  & 18.224 & 0.586 & 17.558 & 0.430 & 16.600 & 0.271 & RRc \\
121 & 17:50:12.18 & -37:02:59.5 & 0.83748  & 18.428 & 0.715 & 17.522 & 0.637 & 16.381 & 0.360 & RRab \\
122 & 17:50:11.78 & -37:03:04.8 & 0.74270  & 18.341 & 1.141 & 17.531 & 0.838 & 16.392 & 0.516 & RRab \\
123 & 17:50:11.44 & -37:03:06.9 & 0.33566  & 18.212 & 0.655 & 17.532 & 0.507 & 16.612 & 0.334 & RRc \\
124 & 17:50:09.73 & -37:02:46.9 & 0.31588  & 18.273 & 0.667 & 17.631 & 0.459 & 16.717 & 0.370 & RRc \\
125 & 17:50:14.36 & -37:02:44.7 & 0.33679  & 18.258 & 0.681 & 17.569 & 0.502 & 16.659 & 0.357 & RRc \\
126 & 17:50:12.54 & -37:03:12.0 & 20.625   & 16.282 & 1.771 & 14.997 & 1.110 & 13.402 & 0.923 & P2C \\
127 & 17:50:12.08 & -37:03:12.3 & 19.773   & 16.398 & 1.521 & 15.048 & 1.057 & 13.441 & 0.869 & P2C \\
128 & 17:50:11.80 & -37:02:59.1 & 13.519   & 16.475 & 0.659 & 15.257 & 0.500 & 13.795 & 0.370 & P2C \\
129 & 17:50:12.87 & -37:03:18.1 & 17.832   & 16.395 & 1.317 & 15.128 & 0.918 & 13.610 & 0.774 & P2C \\
130 & 17:50:14.43 & -37:03:04.8 & 48.90    & 17.412 & 0.799 & 15.500 & 0.628 & 13.364 & 0.406 & LPV \\
131 & 17:50:15.52 & -37:03:07.4 & 122.8    & 17.848 & 1.841 & 15.914 & 1.667 & 12.959 & 1.012 & LPV \\
132 & 17:50:12.87 & -37:03:08.6 & 2.54737  & 17.218 & 1.623 & 16.478 & 0.900 & 15.241 & 0.784 & P2C \\
133 & 17:50:13.98 & -37:02:57.0 & 122.9    & 18.401 & 0.462 & 16.371 & 0.520 & 12.973 & 0.129 & LPV \\
134 & 17:50:15.37 & -37:03:19.5 & 128.9    & 18.655 & 0.869 & 16.731 & 1.088 & 12.980 & 0.556 & LPV \\
135 & 17:50:12.61 & -37:02:38.0 & 162.6    & 18.295 & 4.074 & 16.574 & 3.919 & 13.777 & 2.786 & LPV \\
136 & 17:50:12.69 & -37:03:16.2 & 0.80573  & 18.235 & 0.854 & 17.360 & 0.580 & 16.158 & 0.273 & RRab \\
137 & 17:50:11.85 & -37:03:03.0 & 51.2     & 18.524 & 0.446 & 16.792 & 0.593 & 13.148 & 0.296 & LPV \\
138 & 17:50:14.00 & -37:03:07.7 & 0.80199  & 18.603 & 0.660 & 17.707 & 0.424 & 16.520 & 0.181 & RRab \\
139 & 17:50:13.58 & -37:03:16.2 & 249.1    & 17.792 & 8.083 & 16.009 & 5.819 & 12.982 & 1.716 & LPV \\
140 & 17:50:10.80 & -37:03:05.9 & 0.35181  & 18.244 & 0.421 & 17.539 & 0.372 & 16.643 & 0.277 & RRc; SV15 \\
141 & 17:50:13.98 & -37:03:16.2 & 0.84475  & 18.417 & 0.451 & 17.536 & 0.404 & 16.329 & 0.296 & RRab \\
142 & 17:50:13.83 & -37:02:49.7 & 0.88400  & 18.422 & 0.407 & 17.523 & 0.349 & 16.262 & 0.194 & RRab \\
143 & 17:50:13.75 & -37:02:47.8 & 0.86279  & 18.425 & 0.664 & 17.532 & 0.487 & 16.305 & 0.284 & RRab \\
144 & 17:50:11.34 & -37:02:59.6 & 70.6:    & 18.017 & 0.485 & 16.103 & 0.377 & 13.074 & 0.130 & LPV \\
145 & 17:50:12.72 & -37:03:09.6 & 0.55581  & 18.064 & 0.505 & 17.372 & 0.355 & 16.248 & 0.252 & RRc \\
\enddata 
\end{deluxetable}

\begin{deluxetable}{c|cccc|cccc}
\tablewidth{0pc}
\tablecaption{Comparison of NGC~6441 Variable Stars in the HST and Ground-based 
Surveys \label{tbl-6}} 
\tablehead{
\colhead{ID} & \colhead{$\langle V \rangle_{\rm HST}$} & 
\colhead{$A_{V,{\rm HST}}$} & \colhead{$\langle B \rangle_{\rm HST}$} & 
\colhead{$A_{B,{\rm HST}}$} & \colhead{$\langle V \rangle_{\rm GB}$} & 
\colhead{$A_{V,{\rm GB}}$} & \colhead{$\langle B \rangle_{\rm GB}$} & 
\colhead{$A_{B,{\rm GB}}$} 
          }
\startdata
40 & 17.572 & 1.103 & 18.397 & 1.331 & 17.512 & 1.08 & 18.242 & 1.45 \\ 
55 & 17.562 & 1.054 & 18.333 & 1.507 & 17.523 & 0.97 & 18.194 & 1.25 \\ 
56 & 17.410 & 0.776 & 18.191 & 1.159 & 16.497 & \nodata & 17.595 & 0.64 \\ 
57 & 17.575 & 1.046 & 18.325 & 1.524 & 17.313 & 0.95 & 18.161 & 1.25 \\ 
58 & 17.581 & 0.846 & 18.363 & 1.307 & 16.868 & \nodata & 17.705 & 0.70 \\ 
63 & 17.572 & 0.881 & 18.379 & 1.242 & 17.064 & \nodata & 17.802 & 0.78 \\ 
64 & 17.592 & 0.967 & 18.366 & 1.359 & 16.986 & \nodata & 18.274 & 0.95 \\ 
65 & 17.688 & 0.772 & 18.528 & 0.999 & 16.911 & 0.40 & 17.989 & 0.61 \\ 
75 & 17.471 & 0.355 & 18.139 & 0.469 & 17.349 & \nodata & 18.027 & 0.46 \\ 
93 & 17.474 & 0.412 & 18.020 & 0.592 & 17.332 & 0.54 & 18.057 & \nodata \\ 
95 & 17.605 & 0.559 & 18.369 & 0.793 & 17.611 & 0.55 & 18.256 & 0.67 \\ 
102 & 17.526 & 0.486 & 18.139 & 0.763 & 15.834 & \nodata & 17.257 & 0.32 \\ 
\enddata 
\end{deluxetable}

\begin{deluxetable}{cccccccccc}
\tablewidth{0pc}
\tablecaption{NGC~6441 \& NGC~6388 Population~II Cepheid Mean Properties 
\label{tbl-7}} 
\tablehead{
\colhead{Cluster} & \colhead{ID} & \colhead{Period (d)} & \colhead{$\log\,P$} & 
\colhead{$\langle V \rangle$} & \colhead{$\langle B \rangle$} & 
\colhead{$\langle I \rangle$} & \colhead{$M_V$} & \colhead{$M_B$} & 
\colhead{$M_I$} 
          }
\startdata
NGC~6388 & 18 & 2.89 & 0.4609 & 15.62 & 16.56 & \nodata & --0.77 & --0.23 & \nodata \\ 
NGC~6388 & 29 & 1.88 & 0.2742 & \nodata & 16.04 & \nodata & \nodata & --0.75 & \nodata \\ 
NGC~6388 & 36 & 3.10 & 0.4914 & 15.56 & 16.37 & \nodata & --0.83 & --0.42 & \nodata \\ 
NGC~6388 & 37 & 10.0 & 1.0000 & 14.71 & 15.88 & \nodata & --1.68 & --0.91 & \nodata \\ 
\hline 
NGC~6441 & 6 & 21.365 & 1.3297 & 14.89 & 16.12 & 13.23 & --2.16 & --1.44 & --3.18 \\ 
NGC~6441 & 118 & 0.97923 & -0.0091 & 17.20 & 18.05 & 16.06 & 0.15 & 0.49 & --0.35 \\ 
NGC~6441 & 126 & 20.625 & 1.3144 & 15.00 & 16.28 & 13.40 & --2.05 & --1.28 & --3.01 \\ 
NGC~6441 & 127 & 19.773 & 1.2961 & 15.05 & 16.40 & 13.44 & --2.00 & --1.16 & --2.97 \\ 
NGC~6441 & 128 & 13.519 & 1.1309 & 15.26 & 16.48 & 13.80 & --1.79 & --1.08 & --2.62 \\ 
NGC~6441 & 129 & 17.832 & 1.2512 & 15.13 & 16.40 & 13.61 & --1.92 & --1.16 & --2.80 \\ 
NGC~6441 & 132 & 2.54737 & 0.4061 & 16.48 & 17.22 & 15.24 & --0.57 & --0.34 & --1.17 \\ 
\enddata
\end{deluxetable}

\begin{deluxetable}{ccccccccccc}
\tablewidth{0pc}
\tablecaption{Galactic Globular Cluster Population~II Cepheid Mean Properties 
\label{tbl-8}} 
\tablehead{
\colhead{Cluster} & \colhead{ID} & \colhead{Period (d)} & \colhead{$\log\,P$} & 
\colhead{$\langle V \rangle$} & \colhead{$\langle B \rangle$} & 
\colhead{$\langle I \rangle$} & \colhead{$M_V$} & \colhead{$M_B$} & 
\colhead{$M_I$} & \colhead{References} 
          }
\startdata
NGC~2419     & 18        & 1.578524  &   0.1983 & 18.8    & 19.16   & \nodata & --1.2    & --0.94   &  \nodata & PR77 \\ 
\hline 
NGC~2808     & 10        & 1.76528   &   0.2468 & \nodata & 15.56   & \nodata &  \nodata & --0.26   &  \nodata & CH89 \\ 
\hline 
Palomar~3    & 4         & 3.402     &   0.532  & 19.28   & \nodata & \nodata & --0.69   &  \nodata &  \nodata & BIC00 \\ 
\hline 
NGC~4372     & 24   & 3.13      &   0.50   & 17.04   & 18.22   & 15.75   &   2.00   &   2.79   &   1.20   & KK93 \\
             & 28   & 2.60      &   0.41   & 17.40   & 18.97   & 15.49   &   2.36   &   3.54   &   0.94   & \\
\hline
$\omega$~Cen & 52   & 0.6604    & --0.1802 & 13.95   & \nodata & \nodata & --0.04   &  \nodata &  \nodata & K97 \\ 
             & 60   & 1.3496    &   0.1302 & 13.63   & \nodata & \nodata & --0.36   &  \nodata &  \nodata & \\ 
             & 61   & 2.2743    &   0.3568 & 13.65   & \nodata & \nodata & --0.34   &  \nodata &  \nodata & \\ 
             & 92   & 1.3461    &   0.1291 & 13.98   & \nodata & \nodata & --0.01   &  \nodata &  \nodata & \\ 
             & O123 & 1.0253    &   0.0109 & 13.54   & \nodata & \nodata & --0.45   &  \nodata &  \nodata & \\
             & O156 & 1.0066    &   0.0029 & 13.32   & \nodata & \nodata & --0.67   &  \nodata &  \nodata & \\
             & O161 & 0.9707    & --0.0129 & 13.85   & \nodata & \nodata & --0.14   &  \nodata &  \nodata & \\ 
 & & & & & & & & & \\ 
             & 92   & 1.345     &   0.129  & 13.92   & 14.49   & \nodata & --0.07   &   0.38   &  \nodata & K68 \\ 
 & & & & & & & & & \\ 
             & 1    & 29.18     &   1.47   & 10.89   & 11.66   & \nodata & --3.10   & --2.45   &  \nodata & DC67 \\ 
             & 29   & 14.73     &   1.17   & 11.82   & 12.77   & \nodata & --2.17   & --1.34   &  \nodata & \\
             & 43   & 1.1568    &   0.0633 & 13.36   & 13.82   & \nodata & --0.63   & --0.29   &  \nodata & \\
             & 48   & 4.474     &   0.651  & 12.69   & 13.32   & \nodata & --1.30   & --0.79   &  \nodata & \\ 
             & 60   & 1.34946   &   0.1302 & 13.45   & 13.84   & \nodata & --0.54   & --0.27   &  \nodata & \\ 
             & 61   & 2.27351   &   0.3567 & 13.40   & 14.03   & \nodata & --0.59   & --0.08   &  \nodata & \\ 
             & 92   & 1.34577   &   0.1290 & 13.93   & 14.46   & \nodata & --0.06   &   0.35   &  \nodata & \\ 
\hline 
M3           & 154  & 15.285    &   1.184  & 12.32   & 12.83   & \nodata & --2.81   & --2.31   &  \nodata & DC67 \\ 
 & & & & & & & & & \\ 
             & S7   & 1.3270    &   0.1229 & \nodata & \nodata & \nodata &  \nodata &  \nodata &  \nodata & SED02 \\ 
\hline 
M5           & 42   & 25.728    &   1.410  & 11.28   & 11.82   & \nodata & --3.19   & --2.68   &  \nodata & K68 \\ 
             & 84   & 26.62     &   1.43   & 11.36   & 11.92   & \nodata & --3.11   & --2.58   &  \nodata & \\ 
& & & & & & & & & \\ 
             & 42   & 25.7325   &   1.4105 & 11.31   & 11.85   & \nodata & --3.16   & --2.65   &  \nodata & DC67 \\ 
             & 84   & 26.4945   &   1.4232 & 11.36   & 11.94   & \nodata & --3.11   & --2.56   &  \nodata & \\ 
\hline 
M80          & 1    & 16.3042   &   1.2123 & 13.42   & 14.19   & \nodata & --2.16   & --1.57   &  \nodata & WHB90 \\ 
\hline 
M13          & 1    & 1.459033  &   0.1641 & 14.086  & \nodata & \nodata & --0.41   & \nodata  &  \nodata & KKP03 \\
             & 2    & 5.110818  &   0.7085 & 13.012  & \nodata & \nodata & --1.48   & \nodata  &  \nodata & \\ 
             & 6    & 2.112918  &   0.3249 & 14.078  & \nodata & \nodata & --0.41   & \nodata  &  \nodata & \\ 
 & & & & & & & & & \\ 
             & 1    & 1.459033  &   0.1641 & 14.07   & 14.32   & \nodata & --0.42   & --0.19   &  \nodata & RR83 \\ 
             & 2    & 5.110818  &   0.7085 & 13.12   & 13.57   & \nodata & --1.37   & --0.94   &  \nodata & \\ 
             & 6    & 2.112918  &   0.3249 & 14.03   & 14.46   & \nodata & --0.46   & --0.05   &  \nodata & \\ 
 & & & & & & & & & \\ 
             & 32   & 21.165    &   1.326  & 14.10   & 14.80   & \nodata & --0.39   &   0.29   &  \nodata & RIR82 \\
 & & & & & & & & & \\  
             & 12   & 5.21753   &   0.7175 & 14.70   & 15.15   & \nodata &   0.21   &   0.64   &  \nodata & RR79 \\ 
 & & & & & & & & & \\ 
             & 1    & 1.459252  &   0.1641 & 14.19   & 14.45   & \nodata & --0.30   & --0.06   &  \nodata & PM77 \\ 
             & 2    & 5.110939  &   0.7085 & 13.10   & 13.52   & \nodata & --1.39   & --0.99   &  \nodata & \\
             & 6    & 2.112867  &   0.3249 & 14.11   & 14.54   & \nodata & --0.38   &   0.03   &  \nodata & \\ 
 & & & & & & & & & \\ 
             & 1    & 1.458997  &   0.1641 & 14.00   & 14.20   & \nodata & --0.49   & --0.31   &  \nodata & D71 \\ 
             & 2    & 5.110939  &   0.7085 & 12.98   & 13.29   & \nodata & --1.51   & --1.22   &  \nodata & \\ 
             & 6    & 2.112867  &   0.3249 & 14.00   & 14.42   & \nodata & --0.49   & --0.09   &  \nodata & \\ 
 & & & & & & & & & \\ 
             & 1    & 1.458981  &   0.1640 & 13.69   & 14.19   & \nodata & --0.80   & --0.32   &  \nodata & DC67 \\ 
             & 2    & 5.11128   &   0.7085 & 13.85   & 14.45   & \nodata & --0.64   & --0.06   &  \nodata & \\ 
             & 6    & 2.112897  &   0.3249 & 12.78   & 13.34   & \nodata & --1.71   & --1.17   &  \nodata & \\ 
\hline 
M12          & 1    & 15.527    &   1.191  & \nodata & 12.60   & \nodata &  \nodata & --1.62   &  \nodata & CHY88 \\
\hline  
NGC~6229     & 8    & 14.840457 &   1.1714 & 15.53   & 16.29   & \nodata & --1.92   & --1.17   &  \nodata & BCV01 \\ 
             & 22   & 15.8373   &   1.1997 & \nodata & \nodata & \nodata &  \nodata &  \nodata &  \nodata & \\
\hline 
M10          & 2    & 18.7226   &   1.2724 & \nodata & \nodata & \nodata &  \nodata &  \nodata &  \nodata & CHW85 \\ 
             & 3    & 7.831     &   0.894  & \nodata & \nodata & \nodata &  \nodata &  \nodata &  \nodata & \\ 
 & & & & & & & & & \\ 
             & 2    & 18.728    &   1.272  & 11.83   & 12.65   & \nodata & --2.26   & --1.72   &  \nodata & K68 \\ 
             & 3    & 7.908     &   0.898  & 12.75   & 13.58   & \nodata & --1.34   & --0.79   &  \nodata & \\ 
 & & & & & & & & & \\ 
             & 2    & 18.7351   &   1.2727 & 13.16   & 13.90   & \nodata & --0.93   & --0.47   &  \nodata & DC67 \\ 
             & 3    & 7.87      &   0.90   & 14.09   & 14.89   & \nodata &   0.00   &   0.52   &  \nodata & \\ 
\hline 
M19          & 1    & 16.92     &   1.23   & \nodata & \nodata & \nodata &  \nodata &  \nodata &  \nodata & CH78 \\ 
             & 2    & 14.139    &   1.150  & \nodata & \nodata & \nodata &  \nodata &  \nodata &  \nodata & \\ 
             & 3    & 16.5      &   1.2    & \nodata & \nodata & \nodata &  \nodata &  \nodata &  \nodata & \\ 
             & 4    & 2.4326    &   0.3861 & \nodata & \nodata & \nodata &  \nodata &  \nodata &  \nodata & \\ 
\hline 
NGC~6284     & 1    & 4.48121   &   0.6514 & \nodata & 15.88   & \nodata &  \nodata & --1.20   &  \nodata & CHW80 \\ 
             & 4    & 2.81873   &   0.4501 & \nodata & 16.04   & \nodata &  \nodata & --1.04   &  \nodata & \\ 
\hline 
M92          & 7    & 1.0614007 &   0.0259 & 14.15   & \nodata & 13.61   & --0.52   &  \nodata & --1.04   & K01 \\
\hline  
M9           & 12   & 1.340204  &   0.1271 & \nodata & 16.62   & \nodata &  \nodata &   0.56   &  \nodata & CIR84 \\ 
\hline 
M14          & 1    & 18.729    &   1.273  & 14.03   & 15.35   & \nodata & --2.69   & --1.97   &  \nodata & WF94 \\ 
             & 2    & 2.794708  &   0.4463 & 15.63   & 16.40   & \nodata & --1.09   & --0.92   &  \nodata & \\ 
             & 7    & 13.6038   &   1.1337 & 14.74   & 16.00   & \nodata & --1.98   & --1.32   &  \nodata & \\ 
             & 17   & 12.091    &   1.082  & 14.73   & 15.95   & \nodata & --1.99   & --1.37   &  \nodata & \\ 
             & 76   & 1.890265  &   0.2765 & 15.80   & 16.60   & \nodata & --0.92   & --0.72   &  \nodata & \\ 
 & & & & & & & & & \\ 
             & 1    & 18.734    &   1.273  & 14.06   & 15.28   & \nodata & --2.66   & --2.04   &  \nodata & DW71 \\ 
             & 2    & 2.794708  &   0.4463 & 15.65   & 16.39   & \nodata & --1.07   & --0.93   &  \nodata & \\ 
             & 7    & 13.603    &   1.134  & 14.80   & 16.02   & \nodata & --1.92   & --1.30   &  \nodata & \\ 
             & 17   & 12.085    &   1.082  & 14.81   & 15.95   & \nodata & --1.91   & --1.37   &  \nodata & \\ 
             & 76   & 1.89003   &   0.2765 & 15.84   & 16.60   & \nodata & --0.88   & --0.72   &  \nodata & \\
\hline  
M28          & 4    & 13.462    &   1.129  & \nodata & 14.21   & \nodata &  \nodata & --1.17   &  \nodata & WB90 \\ 
             & 17   & 45.9      &   1.7    & 11.9:   & 13.5:   & \nodata & --3.1:   & --1.9:   &  \nodata & \\ 
 & & & & & & & & & \\ 
             & 4    & 13.458    &   1.129  & \nodata & 14.25   & \nodata &  \nodata & --1.13   &  \nodata & WH84 \\ 
             & 17   & 46.1      &   1.7    & \nodata & 13.5    & \nodata &  \nodata & --1.9    &  \nodata & \\ 
             & 21   & 29.93     &   1.48   & \nodata & 14.75   & \nodata &  \nodata & --0.63   &  \nodata & \\ 
             & 22   & 0.99538   &  -0.0020 & \nodata & 15.75   & \nodata &  \nodata &   0.37   &  \nodata & \\ 
\hline 
M22          & 11   & 1.69050   &   0.2280 & \nodata & \nodata & \nodata &  \nodata &  \nodata &  \nodata & WH78 \\ 
\hline 
M54          & 1    & 1.34769   &   0.1296 & 17.257  & \nodata & 16.589  & --0.368  &  \nodata & --0.849  & LS00 \\ 
             & 2    & 1.09461   &   0.0393 & 17.292  & \nodata & 16.738  & --0.333  &  \nodata & --0.700  & \\ 
\hline 
NGC~6752     & 1    & 1.3782    &   0.1393 & 12.97   & 13.37   & \nodata & --0.18   &   0.18   &  \nodata & L74 \\ 
\hline 
M56          & 1    & 1.510019  &   0.1790 & 15.46   & 16.01   & \nodata & --0.21   &   0.14   &  \nodata & WH85 \\ 
             & 6    & 45.00     &   1.65   & 12.9:   & 13.7:   & \nodata & --2.8:   & --2.2:   &  \nodata & \\ 
\hline 
M15          & 142  & 1.280     &   0.107  & \nodata & 15.66   & \nodata &  \nodata &   0.17   &  \nodata & B98 \\ 
 & & & & & & & & & \\ 
             & 1    & 1.437712  &   0.1577 & 14.996  & \nodata & \nodata & --0.394  &  \nodata &  \nodata & SS95 \\ 
 & & & & & & & & & \\ 
             & 1    & 1.44      &   0.16   & 14.89   & 15.23   & \nodata & --0.50   & --0.26   &  \nodata & H85 \\ 
             & 72   & 1.14      &   0.06   & 15.08   & 15.55   & \nodata & --0.31   &   0.06   &  \nodata & \\ 
             & 86   & 17.11     &   1.23   & 13.7:   & \nodata & \nodata & --1.7:   &  \nodata &  \nodata & \\ 
 & & & & & & & & & \\ 
             & 1    & 1.437395  &   0.1576 & 14.89   & 15.23   & \nodata & --0.50   & --0.26   &  \nodata & DC67 \\ 
\hline 
M2           & 1    & 15.5647   &   1.1921 & 13.36   & 13.97   & \nodata & --2.15   & --1.60   &  \nodata & D69 \\ 
             & 5    & 17.557    &   1.244  & 13.28   & 13.89   & \nodata & --2.23   & --1.68   &  \nodata & \\ 
             & 6    & 19.299    &   1.286  & 13.14   & 13.74   & \nodata & --2.37   & --1.83   &  \nodata & \\ 
             & 11   & 33.5      &   1.5    & 12.11   & 12.67   & \nodata & --3.40   & --2.90   &  \nodata & \\ 
 & & & & & & & & & \\ 
             & 1    & 15.5647   &   1.1921 & 13.39   & 13.99   & \nodata & --2.12   & --1.58   &  \nodata & DC67 \\ 
             & 5    & 17.557    &   1.244  & 13.24   & 13.86   & \nodata & --2.27   & --1.71   &  \nodata & \\ 
             & 6    & 19.299    &   1.286  & 13.06   & 13.70   & \nodata & --2.45   & --1.87   &  \nodata & \\ 
             & 11   & 33.475    &   1.525  & 12.24   & 12.90   & \nodata & --3.27   & --2.67   &  \nodata & \\ 
\hline 
NGC~7492     & 4    & 17.9      &   1.3    & 14.21   & 15.81   & \nodata & --2.86   & --1.26   &  \nodata & B68 \\ 
\enddata 
\tablecomments{PR77$=$Pinto \& Rosino (1977); CH89$=$Clement \& Hazen (1989); 
BIC00$=$Borissova, Ivanov, \& Catelan (2000); KK93$=$Kaluzny \& Krzeminski (1993); 
DC67$=$Dickens \& Carey (1967); SED02$=$Strader, Everitt, \& Danford (2002); 
K97$=$Kaluzny et al.\ (1997); K68$=$Kwee (1968); WHB90$=$Wehlau, Hogg, \& 
Butterworth (1990); Kopacki, Ko{\l}aczkowski, \& Pigulski (2003); RR83$=$Russeva 
\& Russev (1983); RIR82$=$Russeva, Ilieva, \& 
Russev (1982); RR79$=$Russeva \& Russev (1979); PM77$=$Pike \& Meston (1977); 
D71$=$Demers (1971); CHY88$=$Clement, Hogg, \& Yee (1988); Borissova, Catelan, 
\& Valchev (2001); CHW85$=$Clement, Hogg, \& Wells (1985); CH78$=$Clement \& Hogg 
(1978); CHW80$=$Clement, Hogg, \& Wells (1980); K01$=$Kopacki (2001); 
CIR84$=$Clement, Ip, \& Robert (1984); WF94$=$Wehlau \& Froelich (1994); 
DW71$=$Demers \& Wehlau (1971); WB90$=$Wehlau \& Butterworth (1990); 
WH84$=$Wehlau \& Hogg (1984); WH78$=$Wehlau \& Hogg (1978); LS00$=$Layden \& 
Sarajedini (2000); L74$=$Lee (1974); WH85$=$Wehlau \& Hogg (1985); B98$=$Butler 
et al.\ (1998); SS95$=$Silbermann \& Smith (1995); H85$=$Harris (1985); 
D69$=$Demers (1969); B68$=$Barnes (1968)}
\end{deluxetable}

\end{document}